\documentclass[a4paper]{article} 
\usepackage{graphicx}  
\usepackage{graphics} 
\newcommand{\dfr}{d\raise0.3ex\hbox{\kern-0.5ex\char"013 }} 
\newcommand{\lapla}{\bigtriangleup}

\begin{document}

\title{\textbf{Gravitational structure formation in scale relativity}\thanks{Published in: Special Issue of Chaos, Solitons and Fractals on ``New Cosmology" 2003, 16, 565 - 595}}
\author{D. Da Rocha  and  L. Nottale\\{\small CNRS, LUTH, Observatoire de Paris-Meudon,} \\{\small F-92195 Meudon Cedex, France}}
\date{}
\maketitle

\abstract{In the framework of the theory of scale relativity, we suggest a solution to the cosmological problem of the formation and evolution of gravitational structures on many scales. This approach is based on the giving up of the hypothesis of differentiability of space-time coordinates. As a consequence of this generalization, space-time is not only curved, but also fractal. In analogy with Einstein's general relativistic methods, we describe the effects of space fractality on motion by the construction of a covariant derivative. The principle of equivalence allows us to write the equation of dynamics as a geodesics equation that takes the form of the equation of free Galilean motion. Then, after a change of variables, this equation can be integrated in terms of a gravitational Schr\"odinger equation that involves a new fundamental gravitational coupling constant, $\alpha_{g} = w_{0}/c$. Its solutions give probability densities that quantitatively describe precise morphologies in the position space and in the velocity space. Finally the theoretical predictions are successfully checked by a comparison with observational data: we find that matter is self-organized in accordance with the solutions of the gravitational Schr\"odinger equation on the basis of the universal constant $w_{0}=144.7 \pm 0.7$ km/s (and its multiples and sub-multiples), from the scale of our Earth and the Solar System to large scale structures of the Universe.}

\section {Introduction}

One of the main, still open, problems of today's cosmology is that of the formation and evolution of
gravitational structures. In a recent paper \cite{silk01}, J. Silk wrote: ``Galaxy formation theory
is not in a very satisfactory state. This stems ultimately from our lack of any fundamental
understanding of star formation. There is no robust theory for the detailed properties of galaxies."
The same can be said of the formation of planetary systems, as now demonstrated by the discovery of
extrasolar planetary systems with fundamental properties that were totally unexpected from the
standard model of formation. Moreover, at the scale of galaxies and at extragalactic scales, this
question is strongly interconnected with that of ``dark matter". Namely, the existence of large
quantities of unseen matter is a necessary ingredient in the standard approach, since in its absence
the formation of galaxies would be totally impossible. However, while the anomalous dynamical
effects (flat rotation curves of spiral galaxies, velocity dispersion of galaxy clusters, etc...)
and gravitational lensing effects that the dark matter hypothesis attempts to explain are firmly
established, the dark matter itself escapes any detection.

The aim of the present paper is to suggest possible solutions to these problems in the framework of
the theory of scale relativity and fractal space-time. Recall that the introduction of a fractal
and/or Cantorian space-time have been suggested in different contexts and using different tools by
Ord \cite{GO83}, one of us \cite{NS84,IJ1}, El Naschie \cite{EN92} and now many other authors. As we
shall see, this approach provides a solution for both the formation problem (this paper) and the
anomalous effects (joint paper \cite{newcosmuniv}) without needing any additional unseen matter.
Moreover, it allows one to understand the morphogenesis of several structures at all scales and to
theoretically predict the existence of new relations and constraints, that are now successfully
checked from an analysis of the astrophysical data.

After having briefly recalled the foundations of the theory, we apply it more specifically to a
generalized theory of gravitation in which, beyond some scale relative to the system under
consideration, space-time becomes not only curved but also fractal. The induced effects on motion
(in standard space) of the internal  fractal structures (in scale space), are to transform classical
mechanics into a quantum-like mechanics. Then we give the fundamental solutions of the macroscopic
quantum equations, which are adapted to a large class of astrophysical situations (central
potential, constant density, halos). 

Finally the main body of the paper aims at giving for the first time a large panorama of the various
predicted effects and of quantitative and statistically significant verifications in astrophysical
data. We describe  structures, self-organized in terms of the same gravitational coupling constant,
ranging from the scale of our Earth, the Solar System and extra-solar planetary systems, stars
forming zones, galaxies and clusters of galaxies, to large scale structures of the universe, with a
special emphasis about planetary nebulae and our Local Group of Galaxies. It is clear that, due to
the large number and the diversity of the various effects, we cannot enter in the details about the
nature of the observations and the data analysis: several of these results have already been
published in specialized papers that we quote, while publications are in preparation concerning
those which are presented here for the first time. 

\section {Theory}

The construction of the theory is based on the giving up of the Gauss-Riemann hypothesis of local flatness that underlies the building of Einstein's generalized relativity. In other words, we attempt to describe physical laws in a continuous manifold which may be not only curved, but also non-differentiable. The foundation of the theory and its developments has been detailed in several previous papers and books \cite{IJ1,IJ2,liwos,revueFST,quantuniv1,relat}, so that we shall only give here a summary of its tools and its methods.

\subsection{Scale laws}

The first step consists of a description of the internal scale structures in terms of differential
equations written in scale-space. It has indeed been demonstrated
\cite{liwos,revueFST,quantuniv1,benadda00} that continuity and non-differentiability implies
scale-divergence, i.e., fractality, so that non-differentiable coordinates in space-time can be
described in terms of explicitly resolution-dependent variables. 

The simplest possible form for a scale-differential equation describing the dependence of a fractal
coordinate $X$ in terms of resolution $\varepsilon$ is given by a first order, linear,
renormalization-group like equation:
\begin{equation} {\partial X(t,\varepsilon)\over {\partial \ln \varepsilon}}=a - \delta \; X \; ,
\label{eq.4} \end{equation} where $a$ and $\delta$ are independent of $\varepsilon$. Its solution
can be written under the form:  \begin{equation} X(t,\varepsilon) = x(t)\; \left[1+\zeta
(t)\left(\frac{\lambda} {\varepsilon}\right)^{\delta}\right]. \label{eq.5} \end{equation} We
recognize here a fractal asymptotic behavior at small scales with fractal dimension $D=1+\delta$,
which is broken at large scale beyond the transition scale $\lambda$. Moreover, it is easy to check
that the fractal asymptotic behavior comes under the principle of scale relativity, since in a scale
transformation of the resolution, $\varepsilon \rightarrow \varepsilon'$, $\ln X$ transforms
according to the mathematical structure of the Galileo group (but here in scale space).

By differentiating this solution, one obtains the elementary displacement $dX$ as the sum of two
terms: \begin{equation} dX = dx + d\xi. \label{eq.18} \end{equation} Here $d\xi$ represents a
scale-dependent, ``fractal part", and $dx$ a scale-independent, ``classical part" of the full
elementary displacement, which are such that: \begin{equation} dx = v \; dt, \label{eq.19}
\end{equation} \begin{equation} d\xi=\eta \, \sqrt{2 \cal{D}} \; (dt^{2})^{1/2D}, \label{eq.20}
\end{equation} where $<\eta>=0$ and $<\eta^{2}>=1$.

Therefore, while the full variable $X$ is non-differentiable, its ``classical part" $x$ is
differentiable and can therefore be described in terms of standard differentiable calculus. In what
follows we shall consider only the case of a fractal dimension $D=2$, that corresponds to a
Markovian process and that plays the role of a critical fractal dimension \cite{revueFST}.

\subsection{Quantum mechanics induced in space-time}
\subsubsection{Demonstration of Schr\"odinger's equation}
In the minimal theory, two consequences  arise from the non-differentiability and fractality of
space, in addition to the fractality of each individual trajectory (i) which has been established in
the previous section:

(ii) The test-particles can follow an infinity of possible trajectories: this leads one to jump to a
non-deterministic, fluid-like description, in terms of the ``classical part" of the velocity field
of the family of trajectories, $v=v(x(t),t)$. 

(iii) The reflection invariance under the transformation ($dt \leftrightarrow -dt$) is broken as a
consequence of nondifferentiability (see e.g. \cite{quantuniv1}), this leading to a two-valuedness
of the velocity vector. The use of a complex velocity, ${\cal V}= ({v_+}+{v_-})/2 - i
({v_+}-{v_-})/2$ to deal with this two-valuedness can be shown to be a covariant and simplifying
representation \cite{dirac}.

These three effects of nondifferentiability and fractality can be combined to construct a complex
time-derivative operator \cite{liwos} that writes

\begin{equation}
\label{A}
\frac{\dfr}{dt}  =  \frac{\partial}{\partial t} + {\cal V} \cdot \nabla - i \,{\cal D} \, \lapla.
\end{equation}

Using this tool, one can now reformulate the action principle of mechanics. One defines a Lagrange
function ${\cal L}(x,{\cal V},t)$. Since the ``classical part" of the velocity is now complex, the
same is true of the Lagrange function, then of the generalized action ${\cal S}$. A complex
probability amplitude can therefore be introduced as a simple change of variable:

\begin{equation}
\label{3}
\psi=  e^{i {\cal S}/2 m {\cal D}}.
\end{equation}

Finally we can use the equivalence / general covariance principle and write a geodesic equation in
fractal space under the form of the equation of free Galilean motion:

\begin{equation}
\frac{\dfr}{dt}{\cal V}=0.
\end{equation}
After re-expression in terms of $\psi$ and integration, it becomes the free Schr\"odinger equation \cite{liwos}:
\begin{equation}
\label{ABC}
{\cal D}^2 \lapla \psi +  i{\cal D} \frac{\partial}{\partial t} \psi =0.
\end{equation}
This result is supported by a numerical simulation of the underlying fractal process described hereabove performed by Hermann \cite{hermann97}, which allowed him to obtain solutions of the Schr\"odinger equation without explicitly writing it. 

\subsubsection{Generalized theory of gravitation}

We shall not develop in the present contribution the application of this method to standard quantum
mechanics in the microphysical domain (see e.g. \cite{dirac} on this subject). We are interested
here in its macrophysical applications. 

Let us indeed consider the motion of a free particle in a curved space-time whose spatial part is
also fractal beyond some time transition (e.g., the predictability horizon in case of strong chaos)
and/or space transition (e.g., galaxy sizes). Its equation can be written, in the first order
approximation, as a free motion / geodesics equation that combine the general relativistic covariant
derivative (that describes the effects of curvature) and scale-relativistic covariant derivative
(that describes the effects of fractality, Eq. \ref{A}), namely, in the Newtonian limit

\begin{equation}
\frac{\bar{D}}{dt} \;{\cal V}=\frac{\dfr{\cal V}}{dt} +\nabla \left(\frac{\phi}{m}\right) =0,
\end{equation}
where $\phi$ is the Newton potential energy.
Once written in terms of $\psi$ (Eq. \ref{3}), this equation can be integrated in terms of a gravitational Schr\"odinger equation with a potential term:

\begin{equation}
\label{AB}
{\cal D}^2 \lapla \psi +  i{\cal D} \frac{\partial}{\partial t} \psi =\frac{\phi}{2m} \, \psi.
\end{equation}
Since the imaginary part of this equation is the equation of continuity, and basing ourselves on our description of the motion in terms of an infinite family of geodesics, $\rho = \psi \psi^\dagger$ can be interpreted as giving the probability density of the particle positions. For a Kepler potential and in the time-independent (stationary) case, the equation becomes:

\begin{equation}
\label{B}
2{\cal D}^2 \lapla \psi +  (\frac{E}{m}  +  \frac{GM}{r}) \, \psi=0 .
\end{equation}

Even though it takes this Schr\"odinger-like form, this equation is still in essence an equation of gravitation, so that it must keep the fundamental properties it owns in Newton's and Einstein's theories. Namely, it must agree with the equivalence principle \cite{xtrasol,greenberger,agnese},  i.e., it is independent of the mass of the test-particle, while $GM$ provides the natural length-unit of the system under consideration. As a consequence, the parameter ${\cal D}$ takes the form:
\begin{equation}
\label{C}
{\cal D}=\frac{GM}{2 w},
\end{equation}
where $w$ is a fundamental constant that has the dimension of a velocity. The ratio $\alpha_{g}=w/c$ actually plays the role of a macroscopic gravitational coupling constant \cite{agnese,xtrasol2}). As we shall see in what follows, the main result of the theory is that the solutions of this gravitational Schr\"odinger equation are indeed characterized by an universal quantization of velocities in terms of the constant $w=144.7\pm 0.5$ km/s or its multiples or sub-multiples (the precise law of quantization depending on the potential). The universality of this constant is corroborated by its effective intervention in the observed structuration of matter in the Universe on a range of scales reaching 19 orders of magnitude.

Before applying in what follows these fundamental equations of physics to astrophysical problems of formation and evolution of structures, let us conclude this section by recalling that : 
(i) using the same method, Schr\"odinger-like forms have also been obtained for the equation of motion in an electromagnetic field, the Euler and Navier-Stokes equations, the equations of the rotational motion of solids, dissipative systems, and field equations in a simplified case \cite{quantuniv1}; (ii) the quantum rules obtained in this approach are fully demonstrated from first principles, and not a priori set; (iii) the scale relativity method consists of introducing explicitly a scale-space in the physical description. This is a key point for the understanding of our approach. It means that, depending on the scale, either the classical part of the fundamental variables or their fractal part dominates. As a consequence the mathematical tool is quantum at some scales and classical at others, with a transition between the two regimes.

\section {Theoretical predictions}

The generalized gravitational Schr\"odinger equation obtained above can now be used as the motion equation for a large class of systems, namely, all those coming under the three conditions that underlie its demonstration: large number of possible trajectories, fractal dimension 2 of trajectories, and local irreversibility. Actually these conditions amount to a loss of information about angles, position and time.

In general, the equations of evolution are the Schr\"odinger-Newton equation and the classical Poisson equation:
\begin{equation}
\label{Schro}
{\cal D}^2 \lapla \psi + i \,{\cal D} \, \frac{\partial \psi }{\partial t}- \frac{\phi}{2m} \, \psi=0,
\end{equation}
\begin{equation}
\label{Poisson}
\lapla \varphi = 4 \pi G \rho,
\end{equation}
where $\varphi$ is the potential and $\phi=m \varphi$ the potential energy. 

By separating the real and imaginary parts of the Schr\"odinger equation we get respectively a generalized Euler-Newton equation (written here in terms of the Newtonian potential energy $\phi$) and a continuity equation:

\begin{equation}
\label{AA1}
m \, (\frac{\partial}{\partial t} + V \cdot \nabla) V  = -\nabla (\phi+Q),
\end{equation}

\begin{equation}
\label{AA2}
\frac{\partial P}{\partial t} + {\rm div}(P V) = 0,
\end{equation}
This system of equations is equivalent to the classical one used in the standard approach of gravitational structure formation, except for the appearance of an extra potential energy term $Q$ that writes:
\begin{equation}
\label{Q}
Q =-2m{\cal D}^2 \frac{\Delta \sqrt{P}}{\sqrt{P}}. 
\end{equation}

In the case when the particles are assumed to fill the ``orbitals" (for example, the planetesimals in a protoplanetary nebula), the density of matter becomes proportional to the density of probability, $ \rho \propto P=\psi \psi^\dagger$, and the two equations can be combined in terms of a single Hartree equation for matter alone \cite{quantuniv1}:
\begin{equation}
\label{hartree}
\Delta \left(  \frac{{\cal D}^2 \Delta \psi + i {\cal D} \partial \psi/ \partial t}{\psi} \right) - 2 \pi G \rho_0 |\psi|^2 = 0.
\end{equation}
Another situation occurs when the number of bodies is small. They follow at random one among the possible trajectories, so that $P=\psi \psi^\dagger$ is nothing else than a probability density, while space remains essentially empty. This case will be studied in more detail in the joint paper \cite{newcosmuniv}: we suggest that it allows to explain the effects that have up to now been attributed to unseen ``dark matter".

Now, as a first step, we shall mainly study in what follows the simplified case of a potential  which is assumed to be globally unaffected by the structures that it contributes to form. Typical examples are the two-body problem (planetary systems in the Kepler potential of the star, binary systems in terms of reduced mass, ..), cosmology (particles embedded into a uniform density background), ejection processes...).

\subsection{Keplerian potential}

Let us first study the general Keplerian problem. The classical potential $\varphi= -GM/r$ can be inserted in the Schr\"odinger equation:

\begin{equation}
\label{keplertemps}
{\cal D}^{2}\lapla \psi + i{\cal D}\frac{\partial \psi}{\partial t} +
\frac{G\,M}{2r}\psi = 0
\end{equation}
We look for wave functions of the form $\psi =\psi(\textbf{r}) \times \exp(-iEt / 2m{\cal D})$. Making the two substitutions $\hbar/2m \rightarrow {\cal D}$ and $ e^2 \rightarrow GMm$, where $m$ is the test particle inertial mass), we obtain a quantum-like equation \cite{liwos}, \cite{revueFST} which is similar to the quantum hydrogen atom equation (but it is now independent of inertial mass). Thus, we can use the standard solutions (expressed in terms of Laguerre polynomials). In spherical coordinates, the radial part and the angular part are separable: 
\begin{equation}
\label{ah}
\psi(\textbf{r}) = \psi_{nl \hat{m}}(r,\theta,\phi)=R_{nl}(r) \; Y_{l}^{\hat{m}}(\theta,\phi).
\end{equation}
Moreover, the ratio energy/mass is quantized as:
\begin{equation}
\label{energy}
\frac{E_{n}}{m} = - \frac{G^{2} M^{2}}{ 8 {\cal D}^{2} n^2}= - \frac{1}{2} \, \frac{w_{0}^2}{n^2},
 \qquad \textrm{for n} \in \mathbf{N^{\ast}}
\end{equation}
while the natural length unit is the Bohr radius $a_{0} = 4 {\cal D}^{2}/GM =GM/w_{0}^{2}$.

$\bullet$ {\it Consequences for the radial distance distribution:}
Let us consider particles (e.g., of gas, dust, planetesimals in a protoplanetary disk, etc...) involved in highly chaotic and irreversible motion in a central Kepler potential. At time-scales longer than the predictability horizon, the classical orbital elements such as semi-major axes, eccentricities, inclinations, obliquities etc... are no longer defined. However, the stationary Schr\"odinger equation (\ref{B}) that describes their motion in terms of a probability amplitude $\psi$ does have solutions which are characterized by well-defined and quantized values of conservative quantities (prime integrals) such as energy $E$, angular momentum $L$, etc... Therefore we expect the particles to self-organize themselves in the `orbitals' described by these solutions, then to form objects (e.g. planets) by accretion.

After the end of the formation process, the motion of the objects which have been formed remains given by the same values of the prime integrals, thanks to the conservation theorems, but it is either no longer chaotic, or it is characterized by a far larger chaos time (inverse of the Lyapunov exponent). Then one recovers classical orbital elements (semi-major axis $a$, eccentricity $e$, etc...) linked to the conservative quantities by the classical relations, e.g., $E/m=-GM/2a$, $(L/m)^{2}=GM \, a\, (1-e^2)$,...

The theoretical prediction of the probability distribution of a given orbital element can therefore be obtained by searching for the quantum states of a conserved quantity which is a direct indicator of the observable we want to study. This is achieved by choosing the symmetry of the reference system in a way which is adapted to the observable. For example, the spherical symmetry solutions describe states of fixed $E$, $L^2$ and $L_{z}$. From these solutions we can recover the semi-major axis expectation, but not the eccentricity, since the definition of $L^2$ combines $a$ and $e$. Now the parabolic coordinate solutions describe states of fixed $E$, $L_{z}$ and $A_{z}$, where $A$ is the Runge-Lenz vector, which is a conservative quantity that expresses a dynamical symmetry specific of the Kepler problem (see e.g. \cite{landauMEC}). This vector identifies with the major axis and is directed toward the perihelion while its modulus is the eccentricity itself.

The radial part of the Kepler orbitals in a spherical coodinate system are expressed in terms of  Laguerre polynomials. They depend on two integer quantum numbers $n>0$ and $l=0$ to $n-1$. Their average distance is given by (see e.g. \cite{landauMQ}):
\begin{equation}
\label{rm}
{<r>}_{nl} =\, a_{0}\, \left( \frac{3}{2}n^{2} - \frac{1}{2} l(l+1) \right)
\end{equation}

For the maximal value of the angular momentum ($l=n-1$), the mean distance of the test particle becomes ${<r>}_{n} =\,a_{0}\,(n^{2}+\frac{n}{2})$, while the probability peak lies at $r_{n}^{peak} =\,a_{0} \, n^{2}$. The energy being quantized as given in Eq.~\ref{energy}, the semi-major axes can take the values:
\begin{equation}
a_{n}=\frac{GM}{w_{0}^{2}}\,n^2=a_{0}\,n^2.
\end{equation}
The theoretical expectation for the eccentricity distribution is obtained in parabolic coordinates by taking as $z$ axis the major axis of the orbit. One obtains for the projection of the Runge-Lenz vector on this axis \cite{exoecc}: 
 \begin{equation}
A_{z}=e=\frac{k}{n},
\end{equation}
where the number $k$ is an integer and varies from 0 to $n-1$.

$\bullet$ {\it Consequences for the angular distribution:} The angular part of the wave function is also quantized \cite{revueFST}. In spherical coordinates, the angular momenta is quantized as $(L/m)^2 =4{\cal D} \, l(l+1)$ and its projected component as $L_{z}/m=2 {\cal D} \, \hat{m}$, where $\hat{m}$ is the third quantum number and varies from $-l$ to $l$.
The angular solutions are expressed in terms of the spherical harmonics $Y_{l}^{\hat{m}}(\theta,\phi)$, whose importance for morphogenesis will be pointed out in a forthcoming section (ejection process).

$\bullet$ {\it Consequences for dynamics:}
The momentum solutions are separable in spherical coordinates and expressed by the function: $\Psi _ {n, l, \hat{m}} (\textbf {p}) =F _ {n, l} ( p ) \times Y _ { l }^ { \hat{m} } (\vartheta , \phi) $. The functions $ Y _ { l } ^ {\hat{m} } (\vartheta , \phi)$ are standard spherical harmonics, the $F _ {n, l} $ functions are proportional to the Gegenbauer functions $C _ { N } ^ { \nu } ( \chi) $.  The momentum distribution is given by $|p.F_{n,l}(p)|^{2}$. The mean square value of the observable $\textbf {p}$ is given by \cite{bethe}:
\begin{equation}
<p^{2}>=\int_{0}^{\infty}p^{2}|F_{n,l}|^{2}p^{2} \; dp=\left( \frac{p_{0}}{n} \right)^{2},
\end{equation}
where $p_{0}={\hbar}/{a_{0}}$ is the Bohr momentum. We obtain: 
\begin{equation} 
< p^{2} > =  \left(  \frac {GMm} {2n {\cal D}}  \right) ^ {2}= \left( \frac {m \, w _ {0}} 
{n} \right) ^ {2} \quad \textrm{for}\;v<<c, \quad < v^{2} >= \left(\frac { w _ {0}} {n} \right) ^ {2},
\end{equation}
since $ {\cal D} =GM/2w_{0}$. As we shall see in what follows, the observational data supports the universality of the velocity constant $w_{0}$. One indeed finds that matter is self-organized on a wide range of scales in terms of the value $w_{0}=144.7 \pm 0.7$ km/s for this constant (and its multiples and submultiples) \cite{xtrasol}. An attempt of theoretical prediction of this value has been made in Ref.~\cite{xtrasol2}.

\subsection{Harmonic oscillator}

Let us now consider the gravitational potential given by a uniform mass density $\rho$. The domain of application of this case is, evidently, cosmology in the first place, but this can also apply as an approximation to the interior of extended bodies (stars in galaxies, galaxies in superclusters, ...). Solving for the Poisson equation yields a harmonic oscillator gravitational potential ${\varphi}(r)=(2\pi / 3) \, G{\rho} r^{2}$, and the Schr\"odinger equation becomes:
\begin{equation}
\label{oscill}
{\cal D}^{2} \lapla \psi + i {\cal D} \frac {\partial \psi}{\partial t} -
\frac{\pi}{3} G \rho r^{2} \psi \,=\, 0.
\end{equation}

The stationary solutions \cite{revueFST}, \cite{landauMQ} are expressed in terms of the Hermite polynomials ($H_{n}$):
\begin{equation}
R(x,y,z) \propto \exp(- \frac {r^{2}} {2b^{2}} ) \, H_{n_{x}}(\frac{x}{b})
\, H_{n_{y}}(\frac{y}{b}) \, H_{n_{z}}(\frac{z}{b}),
\end{equation}
where $b = {\cal D}^{1/2} \, (\pi G \rho / 3)^{-1/4} $.
The energy/mass ratio is also quantized as $E_{n}/m \, =\, 4{\cal D} \, \sqrt{\pi G \rho / 3} \, (n+ 3/2)$. The main quantum number $n$ is an addition of the three independent axial quantum numbers $n = n_{x}+n_{y}+n_{z}$.

$\bullet$ {\it Spatial consequences:} The theory allows one to predict that matter will have a tendency to form structures according to the various modes of the quantized 3-dimensional isotropic harmonic oscillator \cite{revueFST} whose dynamical symmetry group is SU(3). Depending on the conservative quantities and their associated quantum numbers $(n_{x},n_{y},n_{z})$,  a simple or multiple (double, chain, trapeze) structure is obtained (see Fig.~\ref{fig:oscillatormode}). As we shall see, this prediction can be checked in astrophysical data, since such morphologies are indeed found in the universe on many scales (star formation zones, compact groups of galaxies, multiple clusters of galaxies). Moreover, quantitative predictions can be made: e.g., the distance separation \cite{revueFST} of the extreme density peaks is given by the approximation $ \Delta r_{max}={b}\,({n}^2+3n)^{1/2} $.

$\bullet$ {\it Dynamical consequences:} In the momentum representation, we can predict a distribution of inter-velocities as $\Delta v_{max}={v_0}\,(n^2+3n)^{1/2} $ with the characteristic velocity $v_0= 2\,{\cal D}^{1/2} (\pi G \rho/3)^{1/4}$. The difference between the extreme velocity peaks is quantized in a quasi-linear way \cite{revueFST}, since it is of the order of $?2v_0$, $ 3v_0$, and $ 4v_0$ for the modes $n = 1, 2, 3$ respectively.

The main conclusion is the prediction that the various cosmological constituents of the universe will be situated at preferential relative positions and move with preferential relative velocities, as described by the various structures implied by the quantization of the isotropic 3D harmonic oscillator.

\subsection{Ejection process}

In many ejection / growth processes (planetary nebulae, supernovae, star formation, ejection of matter from Sun , as due e.g. to the infall of sungrazer comets, etc...), the observed ejection velocity seems to be constant in the first approximation (as a result of the cancellation between various dynamical effects). This particular behavior is consistent with a constant or null effective potential, i.e., it corresponds to the free motion case.

This means that this problem becomes formally equivalent to a scattering process during elastic collisions. Indeed, recall that the collision of particles is described in quantum mechanics in terms of an incoming free particle plane wave and of outcoming free plane and spherical waves. The Schr\"odinger equation of a free particle writes:
\begin{equation}
\label{freeSE}
2m {\cal D}^{2} \lapla \psi(\textbf{r}) + E \, \psi(\textbf{r}) = 0
\end{equation}
where $E\, =\, p^{2}/2m \, =2m {\cal D}^{2} k^{2}$ is the energy of this free particle. 
The stationary solutions of the problem in a spherical coordinate system are $\psi(\textbf{r}) \, =\, R(r) \, Y_{l \hat{m}}(\theta,\phi)$. The equation of radial motion becomes \cite{landauMQ}:
\begin{equation}
R^{''}(r) \,+\, \frac{2}{r} \; R^{'}(r)\,+\,\left(k^{2} 
-\frac{l(l+1)}{r^{2}}\right) \, R(r) = 0.
\end{equation}
We keep the solution corresponding to a flow of central particles (ejection / scattering process),
namely, the divergent spherical waves. The general solution is expressed in terms of the first order
Hankel functions:
\begin{equation}
\label{freeradial}
R(r) \,=\, +iA \, \sqrt{\frac{k \pi}{2r}} \; H_{l+\frac{1}{2}}^{(1)}(kr).
\end{equation}

$\bullet$ {\it Radial consequences:} $[\Phi(r)]^{2}$ represents the spatial probability of presence for a particle ejected in a unit of time. But our aim is to know the evolution of the probability function for distances 
and times higher than the ejection area. The spatial 
probability density for a particle emitted at time ($t_{e}$) in an elementary 
spherical volume writes:
\begin{equation}
dP(r,\theta,\phi,t,t_{e}) =\, [R(r-V_{0} \, (t-t_{e}))]^{2} 
\, \frac{1}{r^{2}} \, r^{2} \, dr \; 
[Y_{l \hat{m}}(\theta,\phi)]^{2}\, \sin \theta) \, \sin \phi \; d\theta \, d\phi.
\end{equation}

$\bullet$ {\it Angular consequences:} In a way similar to the interpretation of the spatial solutions, we shall interpret the angular solutions in terms of a self-organized morphogenesis. Indeed, the matter is expected to be ejected with a higher probability along the angle values given by the peaks of the probability density distribution. One may therefore associate quantized shapes, which can be spherical, plane, bipolar, etc... to each couple of quantum numbers $(l,\hat{m})$, i.e., to the discretized values of energy and angular momentum. These different possibilities will be considered in more detail in the section devoted to Planetary Nebulae.

\section {Observational tests of the theory}
\subsection{Solar System}

\subsubsection{Planetary system formation}

\begin{figure}[!ht]
\begin{center}
\includegraphics[width=7cm]{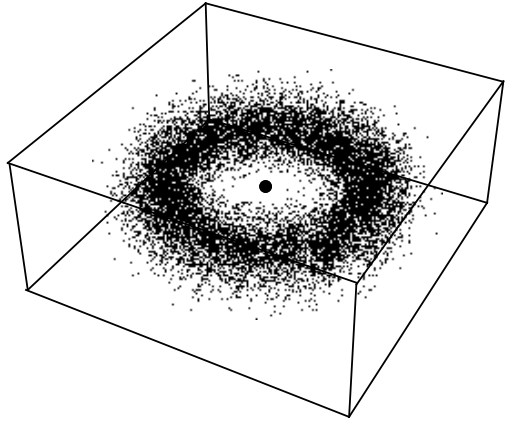}
\caption{Example of solution of the gravitational Schr\"odinger equation for a Kepler potential.}
\label{fig:tore}
\end{center}
\end{figure}

Consider a protoplanetary disk of planetesimals during the formation of a planetary system. The motion of each planetesimal in the central Kepler potential of its star comes under the conditions under which a gravitational Schr\"odinger equation can be written. Therefore we expect the bodies to fill the orbitals (see Fig.~\ref{fig:tore}), then to accrete and to form planets. The final orbital elements of the planets are finally expected to follow the laws that have been described in the previous section, i.e., semi-major axis $\propto n^{2}$, eccentricity $\propto k/n$, etc...

\begin{figure}[!ht]
\begin{center}
\includegraphics[width=7cm]{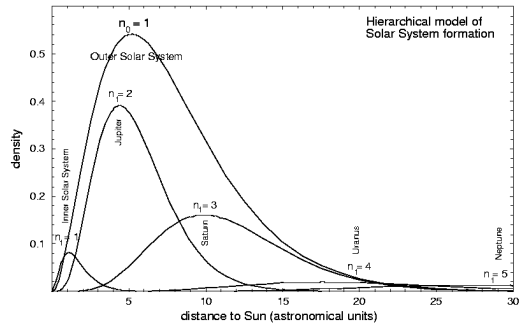}
\caption{Hierarchical model of formation of the Solar System. }
\label{fig:hierarchsosys}
\end{center}
\end{figure}

An important feature of the scale-relativity approach is that it naturally leads to a hierarchy of structures \cite{liwos,xtrasol}. Let us summarize the argument. Consider a system of test-particles (e.g., planetesimals) in the dominant potential of the Sun. Their evolution on large time-scales is governed by Eq.~\ref{B}, in terms of a constant ${\cal D}_{j}=GM/2w_{j}$. The particles then form a disk whose density distribution is given by the solution of the Schr\"odinger equation based on this constant. This distribution can then be fragmented in sub-structures still satisfying Eq.~\ref{B} (since the central potential remains dominant), but with a different constant $w_{j+1}$. We can iterate the reasoning on several hierarchy levels. The matching condition between the orbitals implies $w_{j+1} = k_{j} w_{j}$, with $k_{j}$ integer. Our own Solar System is indeed organized following such a hierarchy on at least 5 levels, from the Sun's radius to the Kuiper belt (see Figs.~\ref{fig:hierarchsosys} and \ref{fig:sosys} and the following sections). In particular, the inner solar system in its whole can be identified as the fundamental level ($n=1$) of the outer solar system, in which Jupiter is in $n=2$, Saturn in $n=3$, etc....

\subsubsection{Planetary orbits}

\begin{figure}[!ht]
\begin{center}
\includegraphics[width=7cm]{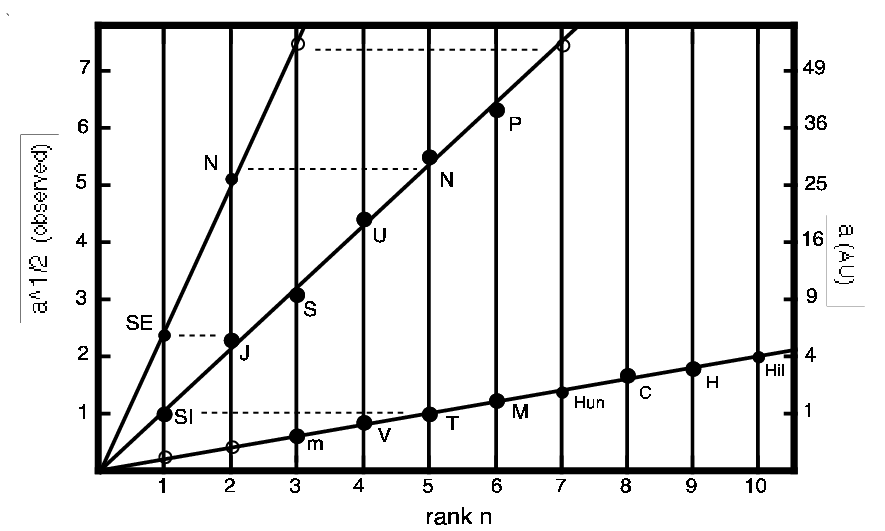}
\caption{Comparison of the observed semi-major axes of planets in the inner and outer solar system with the predicted values, $a_{n}/GM=(n/w)^2$, with $w=144$ km/s in the inner solar system and $w=144/5$ km/s in the outer system. ``SI" stands for the peak of the inner solar system (which corresponds to $n=1$ in the outer system), and ``SE" for the peak of the outer system (Jupiter at $n=2$). Additional peaks are predicted beyond Pluto (see the Kuiper belt section hereafter) and before Mercury (see the section ``Intramercurial structures").}
\label{fig:sosys}
\end{center}
\end{figure}

The application of the scale-relativity approach to the solar system has been given in detail in \cite{liwos,chamonix,sosysNSG}. As expected in the hierarchical model formation, the inner and outer solar systems are both organized in a similar way, in terms of semi-major axes distributed according to a $n^2$ law (see Fig.~\ref{fig:sosys}). For the inner system, the gravitational coupling constant is found to be given by a value $w_{0}=144$ km/s. As we shall see in what follows, this value (or its multiples or submultiples) can be identified on a wide range of astrophysical scales, from the Earth to cosmological scales. 

Moreover, such a $n^2$ law is not adjustable (contrarily to a scaling law of the Titius-Bode type), so that the ranks of the planets are fixed in a constrained way. One finds that Mercury ranks $n=3$, so that the very first result of the theory has been to predict the existence of two additional probable zones for planetary orbits \cite{liwos, chamonix}, at 0.043 AU ($n=1$) and 0.17 AU ($n=2$), for the solar system but also for extrasolar planetary systems. At the time of the prediction, no exoplanet was yet known. As we shall see, this prediction has now received strong support from the discovery of exoplanets and from structures in the very inner solar system. The same is true of the outer solar system beyond Pluto, which can now be checked using the Kuiper belt objects.

\subsubsection{Satellites and rings}
It has been shown by Hermann et al. \cite{hermann98} that the various systems of rings and satellites around the outer giant planets also come under the same $n^2$ law in a statistically significant way. Moreover the planet radii themselves belong to the sequence (as we shall see, the same is true of the Sun in the very inner solar system). Such a result is indeed expected, since a generalized Schr\"odinger equation can also be applied to the problem of the formation of the central bodies, and that matching conditions on the probability amplitude should be written between the interior and exterior solutions (similar to the Cauchy conditions in general relativity). 

In a subsequent study, Antoine \cite{antoine} has solved the Sch\"odinger equation in the case of cylindrical 2D symmetry, and he has compared the solutions to the observed main density peaks of Saturn rings. In this case the energy becomes quantized as $E_{n}=E_{0}/(n+1/2)^2$. The remarkable result is that the ring peaks agree with such a law with the same value $w_{0}=144$ km/s as in the inner solar system: one finds values of $n=$ 6.45, 7.57, 8.51, 9.46 and 13.43, all of them close to  $n+1/2$ as expected. The probability to obtain such a result by chance is smaller than $10^{-3}$.

\subsubsection{Mass distribution of planets}
\begin{figure}[!ht]
\begin{center}
\includegraphics[width=7cm]{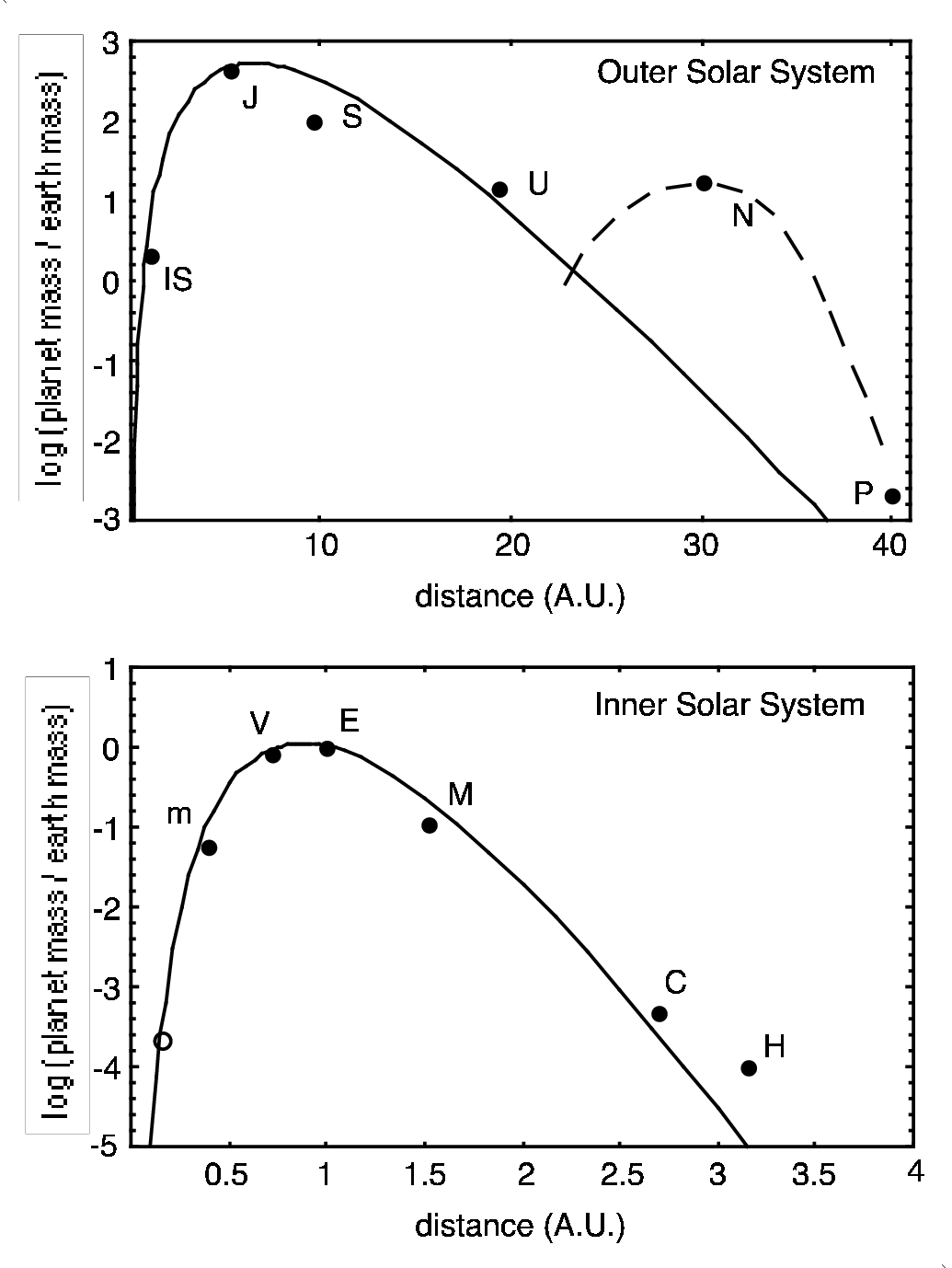}
\caption{Comparison between the observed and theoretically predicted planet masses in the inner and outer solar system. IS stands for the inner solar system as a whole (2 Earth masses), C and H for Ceres and Hygeia, which are the two main mass peaks in the asteroid belt.}
\label{fig:massesplan}
\end{center}
\end{figure}

The above theoretical approach provides a model for the mass distribution of matter in the solar planetary system \cite{chamonix,sosysNSG}. Indeed, as well the observed distribution of the whole system as that of the inner system (which stands globally as the first orbital of the outer one) agree with the predicted law of probability density. We can therefore use the hierarchical model described above to predict the masses of the planets (in units of Jupiter mass, as in runaway models). The result is given in Fig.~\ref{fig:massesplan}.
The distributions of the mass of planets in the inner and outer solar system are in good agreement with such a model. Only the mass of Neptune is much higher than expected. But even this discrepancy is easily explained by the existence of a larger system in which the mass peak of the whole planetary system (i.e. Jupiter) ranks $n=1$, and in which Neptune ranks n = 2 (dashed line in Fig.~\ref{fig:massesplan}; see also  Fig.~\ref{fig:sosys}).

Remark finally that this model will certainly help solving another problem encountered by standard models of planetary formation. The accretion time of planetesimals, though acceptable for earth-like planets, becomes too large for giant planets. In our framework, the initial distribution of planetesimals is no longer flat, but already peaked at about the final value of the planet positions, which should shorten the accretion process.

\subsubsection{Intramercurial structures}

\paragraph{Solar radius}

Preferential distances in the very inner solar system are expected for semi-major axes of 0.043 and 0.17 AU, which correspond to the $n=1$ and $n=2$ probability peaks based on the constant $w_{0}=144$ km/s. More generally one can consider substructures based on a multiple of this constant, $w_1=432=3 \times 144$ km/s. Indeed, the Sun radius is in precise agreement with the peak of the fundamental level of this sequence: namely, one finds $n_{\odot}=0.99$ with $R_{\odot}=0.00465$ AU, that corresponds to a Keplerian velocity of 437.1 km/s. Such a result is not unexpected in the scale-relativity approach. Indeed, the fundamental equation of stellar structure is the Euler equation, which can also be transformed in a Schr\"odinger equation \cite{quantuniv1}, yielding preferential values for star radii. Matching conditions between the probability amplitude that describes the interior matter distribution (the Sun) and the exterior solution (the Solar System) are expected to involve a matching of the positions of the probability peaks. This theoretical question will be developed in a forthcoming work, while data analysis has already demonstrated that known star radii show pronounced probability peaks, in particular for values corresponding to Keplerian velocities of 432 km/s \cite{lefevre}.

\paragraph{Circumsolar dust.}

We therefore predict, on the basis of a constant $w=432$ km/s, probability density peaks lying at $4.09\,R_{\odot}$, $9.2 \,R_{\odot}$, $16.32 \, R_{\odot}$, etc..  This can be checked by studying the density distribution of interplanetary dust.

We have indeed recalled in \cite{sosysNSG} that the possible existence of intramercurial bodies is limited by dynamical constraints (such as the presence of Mercury) and thermodynamical constraints (sublimation). As a consequence asteroids can be found only in the zone 0.1-0.25 AU, but the inner zone 0.005-0.1 AU can yet be checked using the distribution of interplanetary dust particles in the ecliptic plane (originating from comets and asteroids) that produce the F-corona. 

\begin{figure}[!ht]
\begin{center}
\includegraphics[width=11cm]{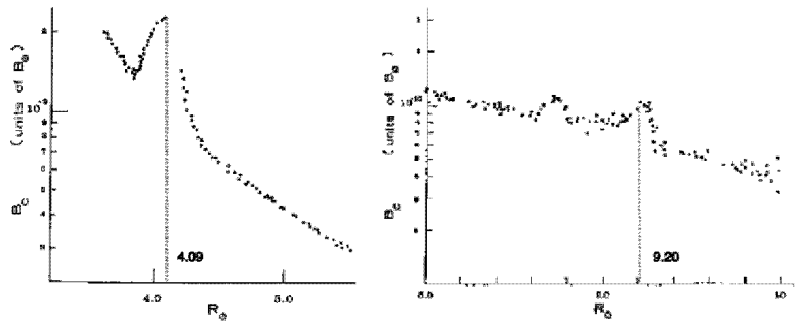}
\caption{IR-dust density observed  during the solar eclipse of January 1967 (adapted from MacQueen \cite{macqueen}). Two density peaks were observed at exactly the predicted distances $4.09\,R_{\odot}$, and $9.20\,R_{\odot}=0.043$ AU, which correspond respectively to Keplerian velocities of $432/2$ and $432/3=144$ km/s.}
\label{fig:IRdust}
\end{center}
\end{figure}

Since 1966, there has been several claims of detection during solar eclipses of IR thermal emission peaks from possible circumsolar dust rings (Peterson \cite{peterson}, MacQueen \cite{macqueen}, Koutchmy \cite{koutchmy}, Mizutani et al \cite{mizutani}, Lena et al \cite{lena}. These structures have been systematically observed at the same distance of the Sun during five eclipses between 1966 and 1983. McQueen finds two radiance peaks at $4.1\, R_{\odot}$, which is equivalent to a Kepler velocity $v=216$ km/s and at $9.2\, R_{\odot} = 0.043$ AU, which corresponds very precisely to a Kepler velocity $v=144$ km/s (see Fig.~\ref{fig:IRdust}). i.e., to the predicted background level of the inner Solar System and of extrasolar planetary systems at the same scale (see herebelow).

However, more sensitive observations during the 1991 solar eclipse \cite{lamy91,hodapp} did not confirm these detections. While Lamy et al. \cite{lamy91} conclude that the previous detections were in error, Hodapp et al. \cite{hodapp} argue that the earliest observations were credible and therefore that the observed structures were transient features, perhaps due to the injection of dust into near-solar space by a Sun-grazing comet. This last interpretation is quite in agreement with the scale-relativity approach. Indeed, the dynamics of the dust particles that are at the origin of the solar F-corona is determined by the Sun gravity, the Poynting-Robertson and corpuscular pressure drag, the radiation pressure force and the effect of sublimation \cite{lamy74}. The combination of these effects leads to chaotic motion with a small predictability time horizon, and would therefore come under the gravitational-Schr\"odinger equation. The dust injected in the circumsolar space would thus accumulate during a finite time into the predicted high probability zones, and finally spiral toward the Sun due to the Poynting-Robertson drag.

Therefore a possible independent test of the theory could consist of new real time IR observations during a forthcoming eclipse taking place just after the passage of large size sungrazer(s).

\paragraph{Vulcanoid belt.}

An attempt to test for the prediction of the theory according to which one or several objects (most probably an asteroid belt) could exist at $0.17\, AU$ from the Sun ($n=2$ for the inner solar system) has been recently performed by Schumacher and Gay \cite{SG01}.

They have tried to detect vulcanoids by analyzing the SOHO/LASCO images using automatic detection. Their conclusion \cite{SG01} is that it is very difficult to detect such objects even if they do exist, because of the high level of noise that remains after cleaning the images for the solar corona. Their result is that there is no object of diameter larger than 60 km around $\approx 0.17 $ AU. This lets open plenty of possibilities, in particular the most probable one, i.e. an asteroid belt made of objects of size smaller than 10 km. Indeed, the prediction of the scale-relativistic planetary model formation (see above) is that a total mass of $10^{-4}$ Earth mass is expected in this zone. It is probable that such a small total mass has not been able to accrete so that the matter remains discretized in terms of planetesimals. Moreover, recent numerical simulations \cite{evans99} have shown that there should exist a stable zone between $0.1$ and $0.2$ AU, and, moreover, that some of the known Earth-crossing asteroids could well come from this zone. 

In situ detection could be possible in the future using e.g. the Solar Orbiter spacecraft, which will reach a distance to the Sun of 0.21 AU. Another suggestion consists of searching for the possible perturbations that such an asteroid belt would induce on the motion of the Aten and Apollo Earth-crossing asteroids, in particular on those which enter the very inner solar system. Six such objects are presently known having perihelion distances smaller than 0.17 (see http://cfa-www.harvard.edu/cfa/ps/mpc.html), among which 1995 CR, of perihelion $q=0.119$ and inclination $i=4.0$ degree, and 2000 BD19 of perihelion $q=0.092$, inclination $i=25.7$ degree, which crosses the ecliptic at the expected belt distance of 0.174 AU.

\subsubsection{Sungrazer comets}
\begin{figure}[!ht]
\begin{center}
\includegraphics[width=7cm]{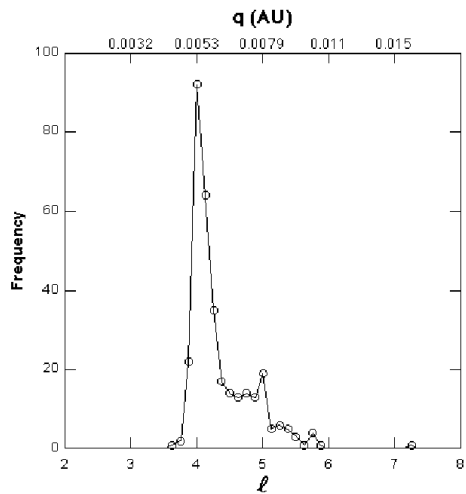}
\caption{Distribution of the perihelions of sungrazer comets and comparison to the scale-relativity prediction (integer values of $l$), according to Ref.~\cite{sungrazers}.}
\label{fig:sungrazers}
\end{center}
\end{figure}

Parabolic comets can be used to check deeper intramercurial structures \cite{sungrazers}. The sungrazers, in particular those recently observed by the SOHO satellite, enter in the very inner solar system. The eccentricity of these objects is very close to $e=1$ because of their quasi-parabolic orbits. Therefore their perihelion distance become a direct indicator of angular momentum. The scale-relativity approach predicts that it should show probability density peaks for perihelion values $q_{l} =(GM/w^2) \, l(l+1)/2 $. The value of $w$ for these objects corresponds to yet a new sub-level of hierarchy, since their velocity may reach 600 km/s. Indeed, as shown in Fig.~\ref{fig:sungrazers}, the perihelion distribution of sungrazers with $q< 0.015$ AU agrees very closely and in a statistically highly significant way with the predicted theoretical values, for $w=1296=9 \times 144$ km/s.  

\subsubsection{Trans-Neptunian structures: Kuiper belt}
\begin{figure}[!ht]
\begin{center}
\includegraphics[width=12cm]{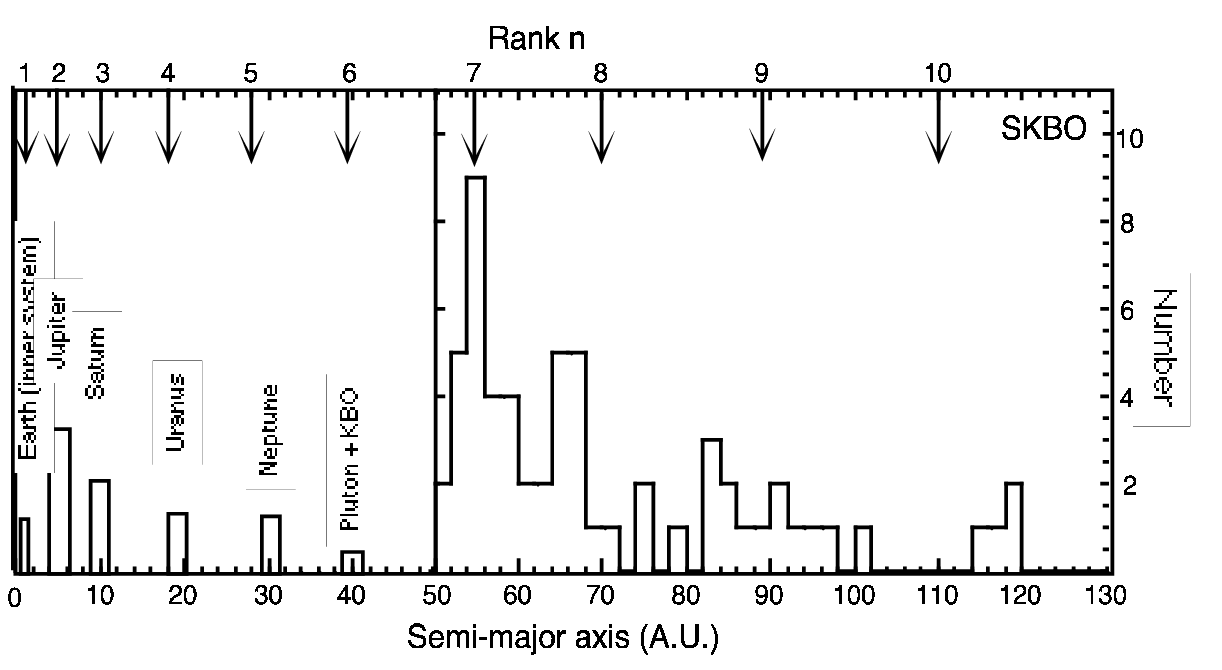}
\caption{Distribution of the semi-major axis of Scattered Kuiper Belt Objects, compared with the theoretical predictions of probability density peaks for the outer solar system (arrows).}
\label{fig:SKBO}
\end{center}
\end{figure}

As recalled above, new predictions have been performed ten years ago in the gravitational Schr\"odinger framework concerning the inner regions of planetary systems, but also the outer regions beyond 40 AU \cite{liwos}. Since, many small bodies have been discovered in the so-called Kuiper belt. For the outer solar system, the $w$ constant is close to $29$ km/s, i.e. 144/5 km/s, where $n=5$ is the rank of the mass peak in the inner solar system, given by the Earth position. This implies a quantization of the semi major axis according to $a_{n}=1.1\,n^{2}$ AU. This law can be used to predict the distribution of Kuiper belt components \cite{sosysNSG}. The distribution of the outer solar system planets and of the recently discovered SKBOs (scattered Kuiper belt objects) is given in Fig.~\ref{fig:SKBO}, and compared to these predictions. A satisfactory agreement is found, in particular concerning the expected trans-Plutonian peak at 55 AU. Moreover, most KBOs are found around 40 AU and they therefore agree with the $n=6$ predicted peak.

\subsubsection{Obliquities and inclinations of planets and satellites}
\begin{figure}[!ht]
\begin{center}
\includegraphics[width=7cm]{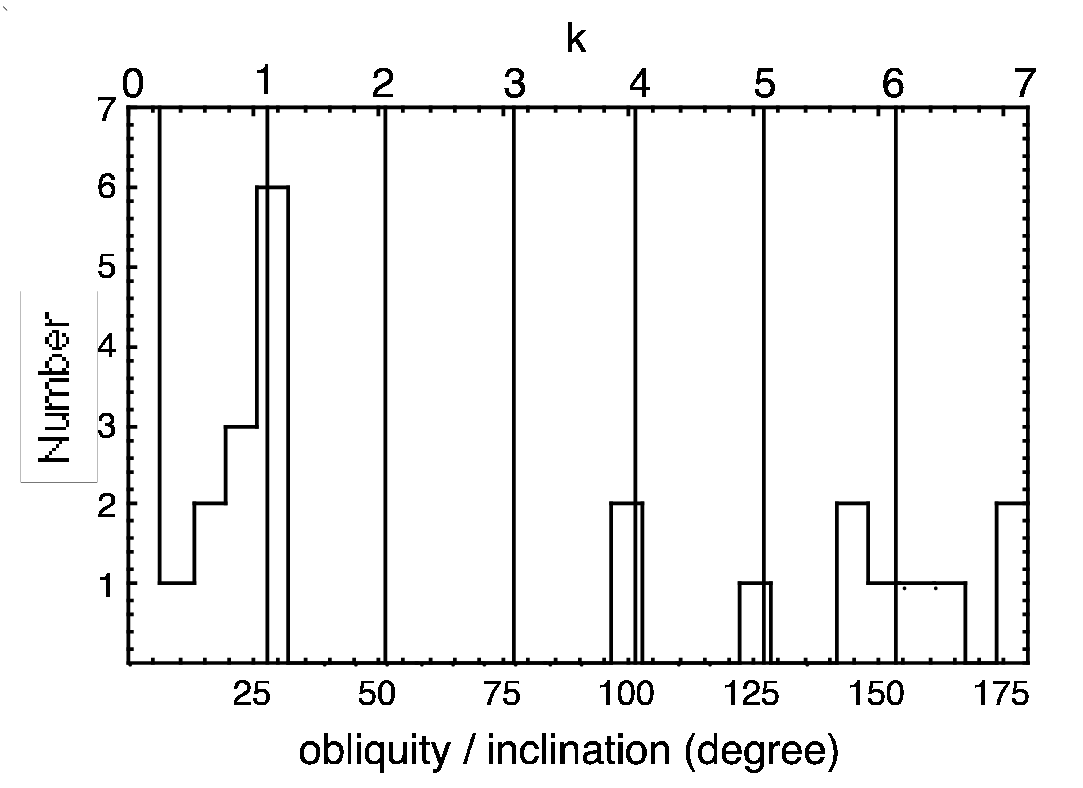}
\caption{Distribution of the obliquities and inclinations of planets and satellites in the Solar System, compared with the scale-relativity prediction (integer values of $k$).}
\label{fig:obliquity}
\end{center}
\end{figure}

The method of transformation of classical dynamics equations in Schr\"odinger equations under the information-loss conditions has also been applied to the equations of the rotational motion of solids \cite{obliq,quantuniv1}. The planetary and satellite obliquities in the solar system are known to be most of the time chaotic and they therefore come under the Schr\"odinger description. In the case of a free rotational motion, one obtains an equation that can also be applied to their inclinations:
\begin{equation}
\frac{d^{2}\psi}{d \theta^{2}} + A^{2}\psi = 0
\end{equation}
Since $\theta$ can vary only between 0 and $\pi$, the expected probability density of angle values is $P(\theta)=a\,\cos^{2}(n\theta)$. Thus the probability peaks are predicted to be quantized as: $\theta_{k}=k \pi / n$, where $n$ is an integer. For the whole solar system, the observed values of obliquities and inclinations agree in a significant way with the predicted distribution associated to $n=7$ (see \cite{obliq} and Fig.~\ref{fig:obliquity}). Note that the earth obliquity itself $(23^{\circ}27^{'})$ and several other bodies of the solar system fall close to the second quantized value $k=1$. The retrograde planets such as Venus, and those which are almost heeled over their orbital plane, such as Uranus, are also accounted for in this model.

\subsubsection{Space debris around Earth}
The predictions of the scale relativity theory are clearly consistent with the observation of many quantized gravitational systems. Now it could be interesting to have not only observational, but also experimental validations of the scale-relativity proposal.

One could therefore suggest a gravitational experiment, consisting of sending test particles in chaotic motion in a Keplerian gravitational potential. Actually such an experiment has already been done, even though it was not in purpose. Indeed, diffusing space debris in orbit around the Earth provides us with exactly this kind of scientific experiment. A large class of these space debris have diffused through collision process so that the information about their initial orbits is lost. Thus, we can apply to these objects the Keplerian solutions allowed by the theory. 

Starting from the best fit value of the fundamental gravitational velocity constant for inner solar system planets and exoplanets (see below), $w_0 = 144.3$ km/s, and with the Earth mass of $5.977 \times10^{24}$ kg, we expect average orbiting distance given by: ${<r>}_{n} =(G M / w_{0}^2) \,(n^{2}+n/2)\,=\,19.15 \times (n^{2}+n/2)$ km.
For $n=18$, we find the Earth radius with a remarkable precision: $GM(n^{2}+n/2)/w_{0}^2=6375$ km, while the equatorial radius of the Earth is $6378.160$ km (the difference between levels amounting to about 700 km). This result, though it needs further theoretical analysis, is expected for the same reasons as the connection of the Sun radius with the Solar System structures (see above section ``Intramercurial structures"). Thus the mean distance of the space debris are predicted to be given by the next probability peaks at $718$ km ($n=19$), $1475$ km ($n=20$), $2269$ km ($n=21$), etc....

\begin{figure}[!ht]
\begin{center}
\includegraphics[width=7cm]{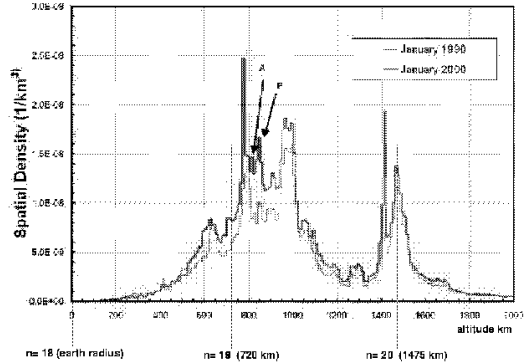}
\caption{Density distribution of the recently observed nearest space debris (adapted from \cite{anz}). Theoretically predicted peaks lie at altitudes of $720$ km and $1475$ km.}
\label{fig:space-debris}
\end{center}
\end{figure}

The available data ($< 2000$ km) \cite{anz} clearly shows two density peaks at $850$ km and $1475$ km (see Fig.~\ref{fig:space-debris}). The second peak is in total accordance with the prediction (however a more complete analysis is still needed to verify that it does not correspond to a predetermined orbit). Concerning the first peak, it is necessary to take into account the dynamical braking of the earth atmosphere. This braking deviates particles to the Earth in a region up to about $700$ km. Moreover, the observed distribution should also be corrected for spikes that correspond to identified debris still orbiting about their original orbit. These corrections should be performed before reaching a firm conclusion about this test of the theory. 

\subsection{Extrasolar Planetary system}

\subsubsection{Semi-major axis}
One of us wrote in a colloquium about ``Chaos and diffusion in Hamiltonian systems", held in February 1994, i.e. more than one year before the discovery by Mayor and Queloz \cite{51Peg} of the first exoplanet around a solar-type star \cite{chamonix}: ``One of the difficulty of theories of the Solar System formation and structures is, up to now, its uniqueness: we do not know whether an observed  ``law" is a peculiar configuration of our own system, or whether it is shared by all planetary systems in the universe. But we can expect such other systems to be discovered in the forthcoming years, and new informations to be obtained about the very distant solar system (Kuiper's belt, Oort cometary cloud...). In this regard our theory is a falsifiable one, since it makes definite predictions about such observations of the near future: observables such as the distribution of eccentricities, mass, angular momentum, the preferred positions of planets and asteroids, or possibly the ratio of distance of the largest gaseous planet and the largest telluric one, are expected in our framework to be universal structures shared by any planetary system."

We have seen in the preceding section how the theoretical prediction made in this text about the Kuiper belt (and other structures in the Solar System) is now being successfully checked. Since now seven years, the discovery of more than 70 exoplanets allows us to put to the test the second prediction, i.e. the universality of planetary structures.

Concerning semi-major axes, the presently observed exoplanets fall in the distance range of the inner solar system. The $n^2$ law of the inner solar system can be tested by considering for each exolanet the variable $4.83\,\sqrt{a/M}$, where $a$ is the planet semi-major axis (in AU), $M$ the star mass (in solar mass unit) and $w_{0}=4.83$ (i.e. 144 km/s in Earth velocity unit). The theoretical expectation is therefore that this variable should cluster around integer values  (without any free parameter and performing no fit of the data). The result for the exoplanets known at the date of writing of the present contribution is given in Fig.~{fig:exoplanet-a} and nicely supports the prediction. The probability to obtain such an agreement between the data and the theoretical prediction is $4 \times 10^{-5}$ (\cite{xtrasol2,exoecc}). 

A particularly remarkable result concerns the $n=1$ and $n=2$ orbitals at 144 and 72 km/s, where no large planet is present in our own system (as expected from its mass distribution determined by the distance of Jupiter to the Sun), but on which a large number of exoplanets have been found. The proximity of these exoplanets (the so-called 51 Peg-like exoplanets) to their star is a puzzle for standard theories of formation, while it was predicted in advance in the scale-relativity framework, moreover in a quantitative way, since these planets orbit preferentially at $a/M$=0.043 and 0.17 AU/M$_{\odot}$.
\begin{figure}[!ht]
\begin{center}
\includegraphics[width=9cm]{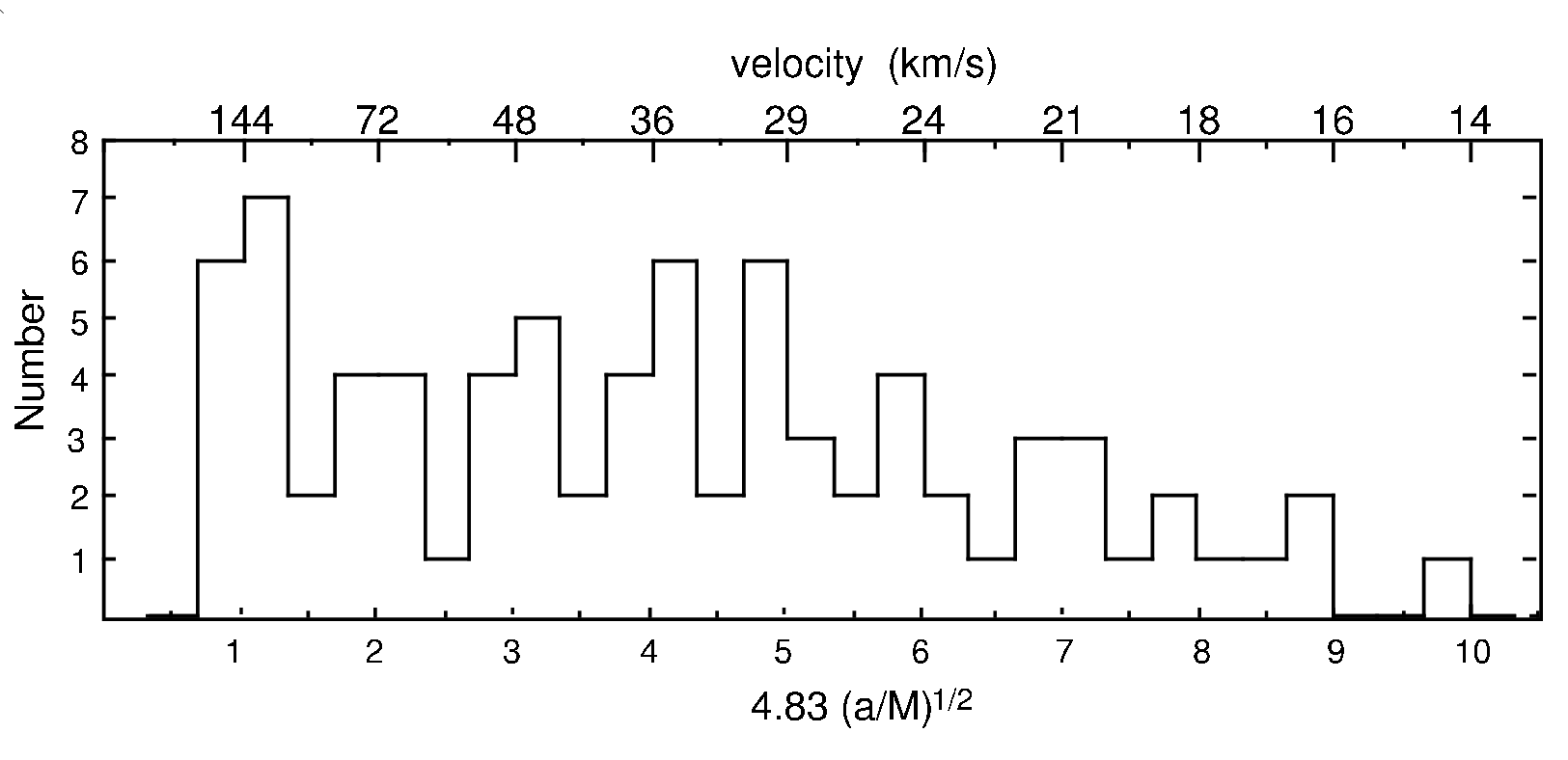}
\caption{Distribution of the semi-major axis of 79 planets (inner solar system and exoplanets around solar-like stars). Mercury, Venus, the Earth and Mars lie respectively at $n=3, 4, 5$ and $6$.}
\label{fig:exoplanet-a}
\end{center}
\end{figure}

\subsubsection{Eccentricity}
\begin{figure}[!ht]
\begin{center}
\includegraphics[width=7cm]{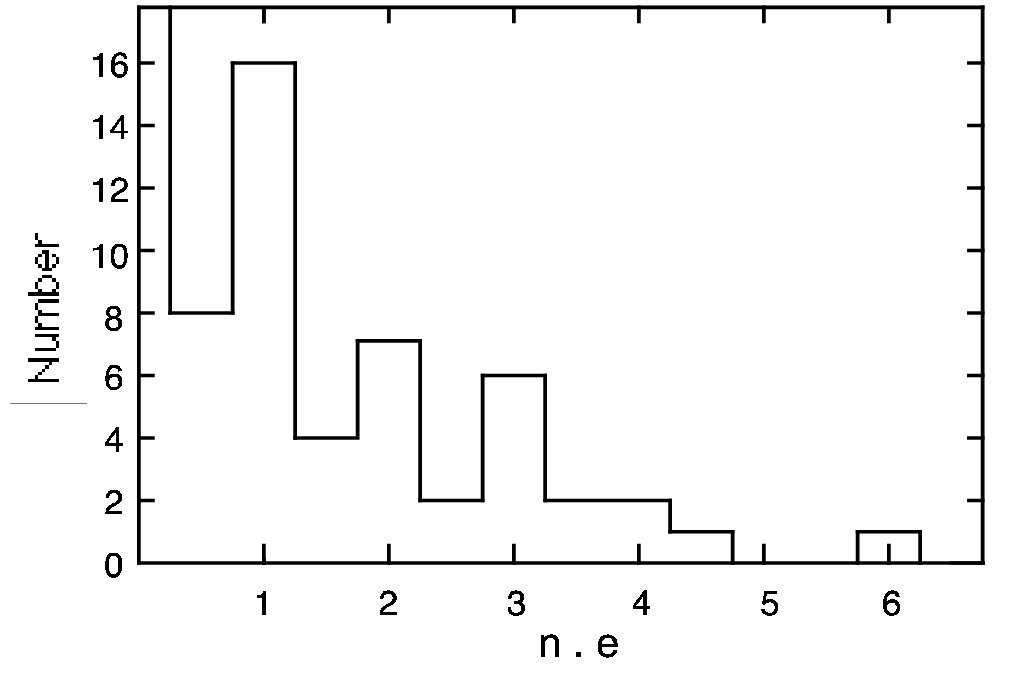}
\caption{Distribution of the eccentricities of 79 planets (inner solar system and exoplanets around solar-like stars). The theory predicts that the product of the eccentricity $e$ by the quantity $\tilde{n}=4.83 (a/M)^{1/2}$ should cluster around integers.}
\label{fig:exoplanet-e}
\end{center}
\end{figure}

The eccentricity distribution of the exoplanets can be studied with regard to the general Keplerian eccentricity solution, $e =k / n$. The eccentricity distribution of the global sample (combining all the exoplanets and the inner solar system bodies) agrees with the predicted quantized distribution around integers $k=n \times e$ (see Fig.~\ref{fig:exoplanet-e}). The associated probability level is $10^{-4}$ \cite{exoecc}. When combining the eccentricity and semi-major axis distribution, the probability to find such a distribution by chance becomes as low as $3 \times 10^{-7}$. 

The discovery of the large eccentricities of exoplanets is the second puzzle posed to standard models of formation. On the contrary, this is an expected result in our framework, since we predict the existence of orbits with eccentricities ranging from $e=0$ to $e=1-1/n$.

\subsubsection{Pulsar exoplanets}
The first discovery of an extrasolar planetary system are the three planets found by Wolszczan around the pulsar PSR B1257+12 \cite{pulsar}. Even though the star is not solar-like, it deserves a special study \cite{xtrasol,pulsar}. Indeed the planets probably result from an accretion disk formed around the very compact star after the supernova explosion. We can therefore expect that the purely gravitational formation process described by the gravitational Schr\"odinger equation be valid with very few perturbations, so that the theoretical predictions would become very precise. Moreover, the compacity of the star suggests that planets be self-organized in terms of a smaller scale than the inner solar system (i.e., of a multiple of $w_{0}=144$ km/s.

\begin{figure}[!ht]
\begin{center}
\includegraphics[width=8cm]{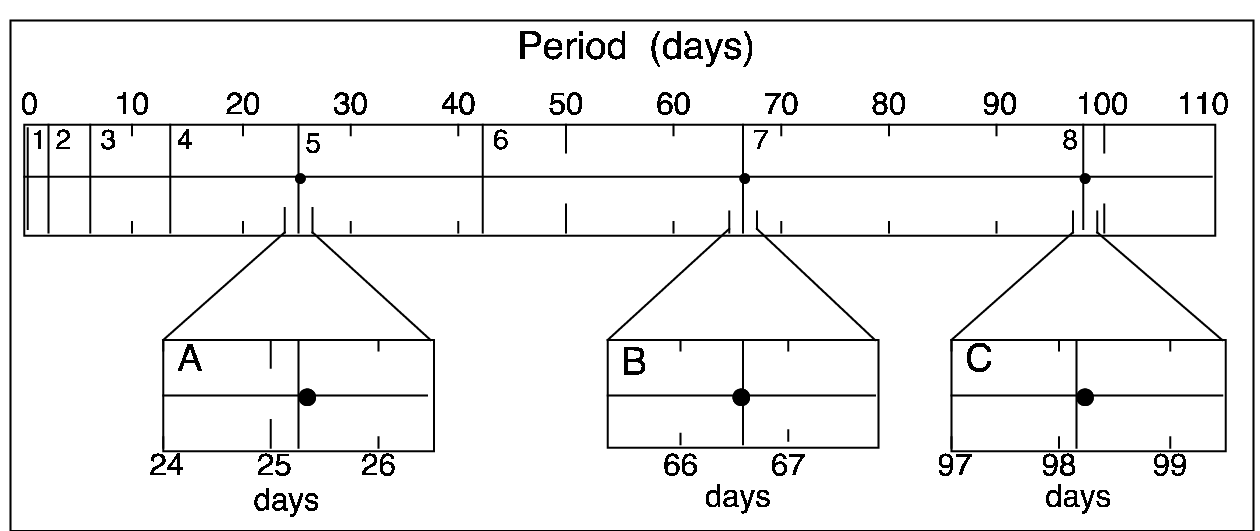}
\caption{Comparison between the observed periods of the three planets observed around the pulsar PSR B1257+12 and the scale-relativity prediction. The agreement between the observed periods and the predicted ones is so precise (the difference is less than 3 hours for periods of several months) that three zooms by a factor of 10 are indicated.}
\label{fig:PSR}
\end{center}
\end{figure}

Both expectations are supported by the data (see Fig.~\ref{fig:PSR}. Indeed, assuming that the planets finally formed at mean distance of the orbitals, we expect the final planets to orbit with periods given by $P_ {n}=(2 \pi G M/ w^{3}) \, (n^{2} + n/2)^{3/2} $, where $M$ is the pulsar mass. Period ratios from this formula (with $n=$ 5, 7 and 8 for the three planets A, B and C) agree with the observed ratios with a remarkable precision of some $10^{-4}$: one obtains $(P_{A}/P_{C})^{1/3}=0.6366$ while $(P_{5}/P_{8})^{1/3}=0.6359$ and $(P_{B}/P_{C})^{1/3}=0.8783$ while $(P_{7}/P_{8})^{1/3}=0.8787$ \cite{pulsar}. Moreover, using the standard pulsar mass $M=1.4 \pm 0.1$ M$_{\odot}$, one obtains for the coupling constant of this system $w=(2.96 \pm 0.07) \times 144$ km/s (i.e. the Keplerian velocity at the Sun radius). 

These results have allowed to predict precisely the possible existence of other periods, in particular short ones at 0.322 days ($n=1$, 1.958 days $n=2$ and 5.96 days ($n=3$). In a recent work, Wolszczan et al. \cite{wolszczan00} have obtained timing data for about 30 successive days. We have analyzed the residuals of these data after substraction of the effects of the three planet (the dispersion of these residual being still larger than the error bars), and we have found a marginal detection (at a significance level of 2.7 $\sigma$) for a period $P=2.2$ days, which is compatible with the $n=2$ prediction. Such a preliminary result clearly needs confirmation using more complete data.

\subsection{Galactic structures}

\subsubsection{Star formation}
\begin{figure}[!ht]
\begin{center}
\includegraphics[width=8cm]{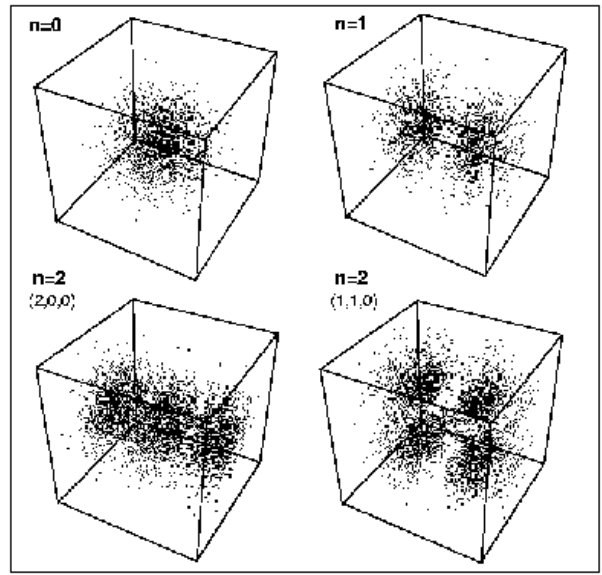}
\caption[1]{The three first modes of the 3-dimensional harmonic oscillator potential, that correspond to the gravitational potential of a background of constant density (the mode $n=2$ decays into two sub-modes). In the scale-relativity approach, the geodesics equation can be integrated in terms of a Schr\"odinger equation, so that structures are formed even in a medium of strictly constant density. Depending on the value of the energy, discretized stationary solutions are found, that describe the formation of one object ($n=0$), two objects ($n=1$), etc... We have simulated these solutions by distributing points according to the probability density. The mode $n=1$ corresponds to the formation of binary objects (stars, galaxies, clusters of galaxies...).}
\label{fig:oscillatormode}
\end{center}
\end{figure}

The new approach allows to bring simple solutions to some still unsolved fundamental problems in the standard theory of star formation. This concerns in particular the morphogenesis of the star forming zones. Indeed, one can as a first approximation describe the interstellar medium from which a star forms in terms of an average constant density, i.e., of a 3-D isotropic harmonic oscillator potential. The solutions of the corresponding Schr\"odinger equation are given in Sect.3.2 and illustrated in Fig.~\ref{fig:oscillatormode}. For an increasing energy, these solutions describe single objects ($n=0$), then binary structures ($n=1$), then 3-chains and trapeze-like structures ($n=2$), 4-objects chains ($n=3$), etc...  Such morphologies are naturally unstable and rapidly evolve just after their formation, since the potential is locally changed by the structuration.

\begin{figure}[!ht]
\begin{center}
\includegraphics[width=8cm]{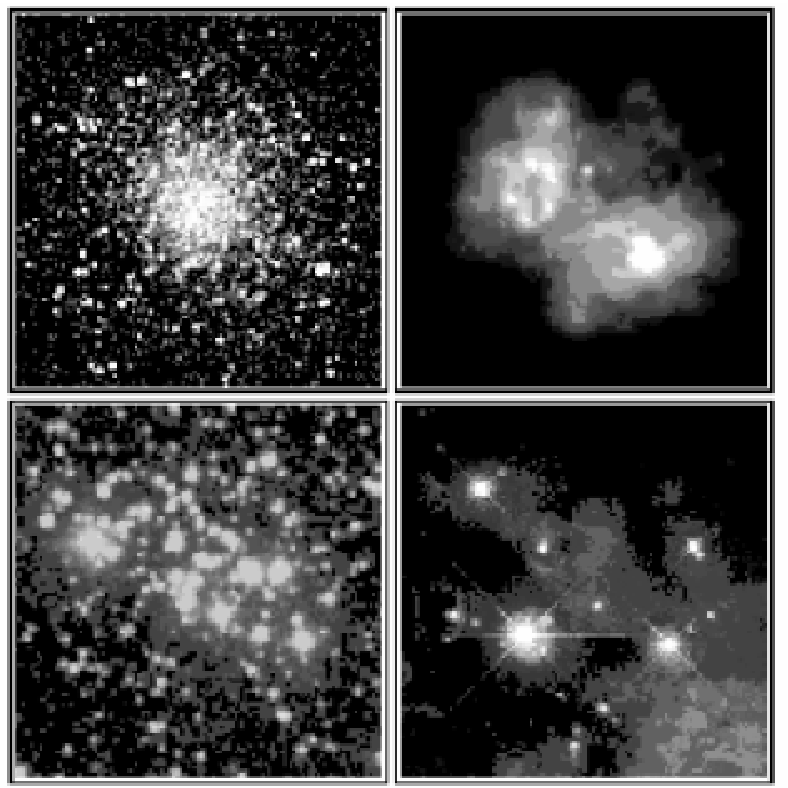}
\caption[1]{Examples of morphologies of star clusters. Up left: the globular cluster M13; up right: the ``butterfly" nebula in M59 \cite{heydari}; down left: the chain cluster NGC 7510; down right: the trapeze in Orion nebula.}
\label{fig:starclusters}
\end{center}
\end{figure}

Now it is remarkable that zones of star formations such as O and B associations are known to show in their central regions double structures (ex. the ``butterfly" in N159 of the Large Magellanic Cloud), chain-like morphologies (ex. NGC 7510), trapeze-like morphologies (ex. the Orion trapeze), etc... (see Fig~\ref{fig:starclusters}). 

Another morphological specificity of star formation at a smaller scale is the presence of disks associated with polar jets. As we have seen in the ``Ejection process" section and will be illustrated hereafter in the ``Planetary Nebula" section (see the cases ($l=2$, $m=0$) and ($l=4$, $m=0$), this is precisely the result obtained when looking for the angular dependence of the solutions of the Schr\"odinger equation, when assuming that the matter and the gas has been preferentially ejected at angles given by the peaks of angular probability density. This vast subject can be only touched upon in this review and will be developed in more detailed in a forthcoming work.

\subsubsection{Binary stars}
A crucial test of the theory consists of verifying that it applies to pure two-body systems. The formation of binary stars remains a puzzle for the standard theory, while more than 60 \% of the stars of our Galaxy are double. Conversely, in the new approach the formation of a double system is obtained very easily, since it corresponds to the solution $n=1$ of the gravitational Schr\"odinger equation in an harmonic oscillator potential, i.e., a uniform density background (while the fundamental solution $n=0$ represents a single spherically symmetric system): see Figs.~\ref{fig:oscillatormode} and \ref{fig:starclusters}.

After its formation, the binary system will evolve according to its local Kepler potential. The binary Keplerian problem is solved, in terms of reduced coordinates, by the same equations as single objects in a central potential. This solution brings informations about the inter-velocities and the inter-distances between stars. The observed velocity is expected to be quantized as $v_{n}=w/n$ with $w$ equal to $144$ km/s or a multiple or submultiple (depending on the scale of the binary star). 

\begin{figure}[!ht]
\begin{center}
\includegraphics[width=7cm]{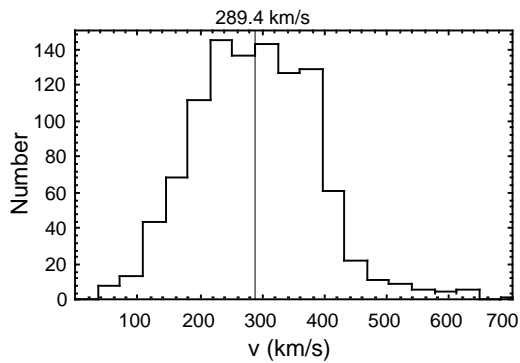}
\caption{Distribution of the intervelocity between binary stars in the Brancewicz catalog of eclipsing binaries. The average velocity is $ 2 \times (144.7 \pm 1.5)$ km/s.}
\label{fig:brancewicz}
\end{center}
\end{figure}

As an example of application, eclipsing binaries are an interesting case of close-by systems, for which we therefore expect the gravitational constant to be a multiple of 144 km/s. It has indeed been found \cite{fractal98,binstar} that the average velocity of the 1048 eclipsing binaries in the Brancewicz and Dvorak catalog of eclipsing binaries \cite{brancewicz} is $w = 289.4 \pm 3.0 = 2 \times (144.7 \pm1.5)$ km/s (see Fig.~\ref{fig:brancewicz}). Moreover, a good fit of their interdistance distribution is given by the probability distribution of the fundamental Kepler orbital; similar results are obtained using several other catalogs of binary stars \cite{binstar}.

\subsubsection{High-velocity clouds}
\begin{figure}[!ht]
\begin{center}
\includegraphics[width=14cm]{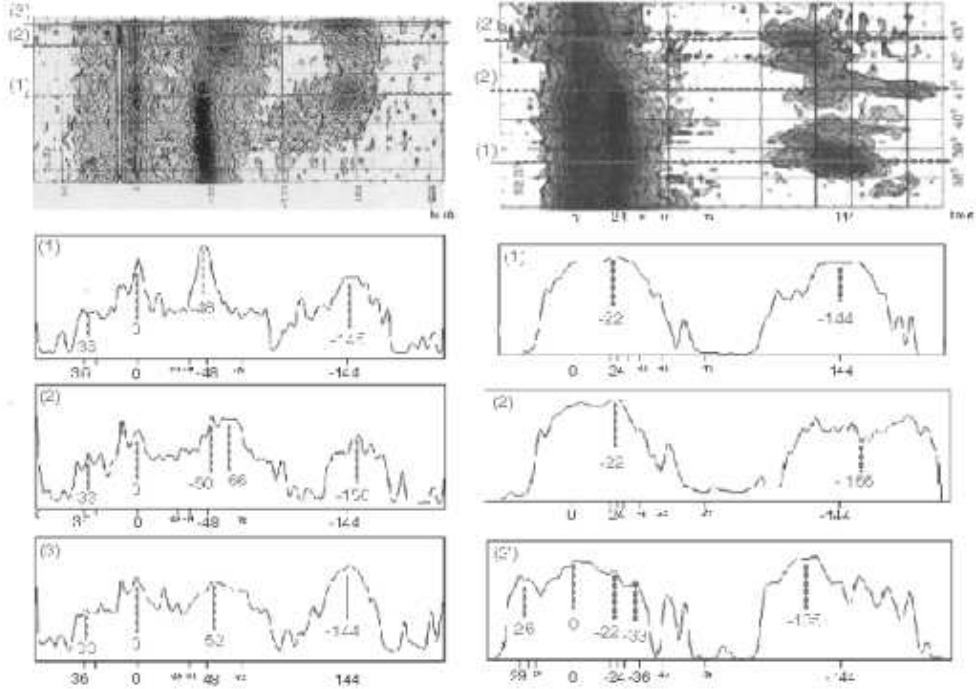}
\end{center}
\caption{Internal structures of HVCs in complex C (centered on $b=54^{\circ}.6$ and $l=92^{\circ}.4$). Details reveal a Keplerian distribution in $144/n$ km.s$^{-1}$. (Adapted from ref.~\cite{pietz}).}
\label{hvc}
\end{figure}

High-velocity clouds (HVCs) are neutral gas clouds with anomalous observed velocities. They are grouped in large structures named complexes. In a recent paper, Woerden et al. \cite{vanW} have obtained more precise values for the distances of HVCs in complex A (chain A) and confirmed their positions in the halo. A particular structure of such typical clouds has been observed by Pietz et al. \cite{pietz} in the complex C. Velocity bridges have been identified, that seem to be in accordance with the expected Keplerian velocity distribution $(144/n$ km.s$^{-1})$. We give two examples of this effect in Fig.~\ref{hvc}, which shows the mass density toward the line of sight in function of the radial velocity (the zero velocity is associated to the HI gas in the Galactic disk). Large bridges at $144$ km.s$^{-1}$ with sub-structures near $\pm 20$ km.s$^{-1}$, $\pm 24$ km.s$^{-1}$, $\pm 36$ km.s$^{-1}$ and $- 48$ km.s$^{-1}$ are clearly apparent in these diagrams.
This distribution can be considered as a signature of a Keplerian interaction between the molecular clouds in the Galactic halo and the Galactic disc.

\subsubsection{Proper motion near the Galactic center}
\begin{figure}[!ht]
\begin{center}
\includegraphics[width=7cm]{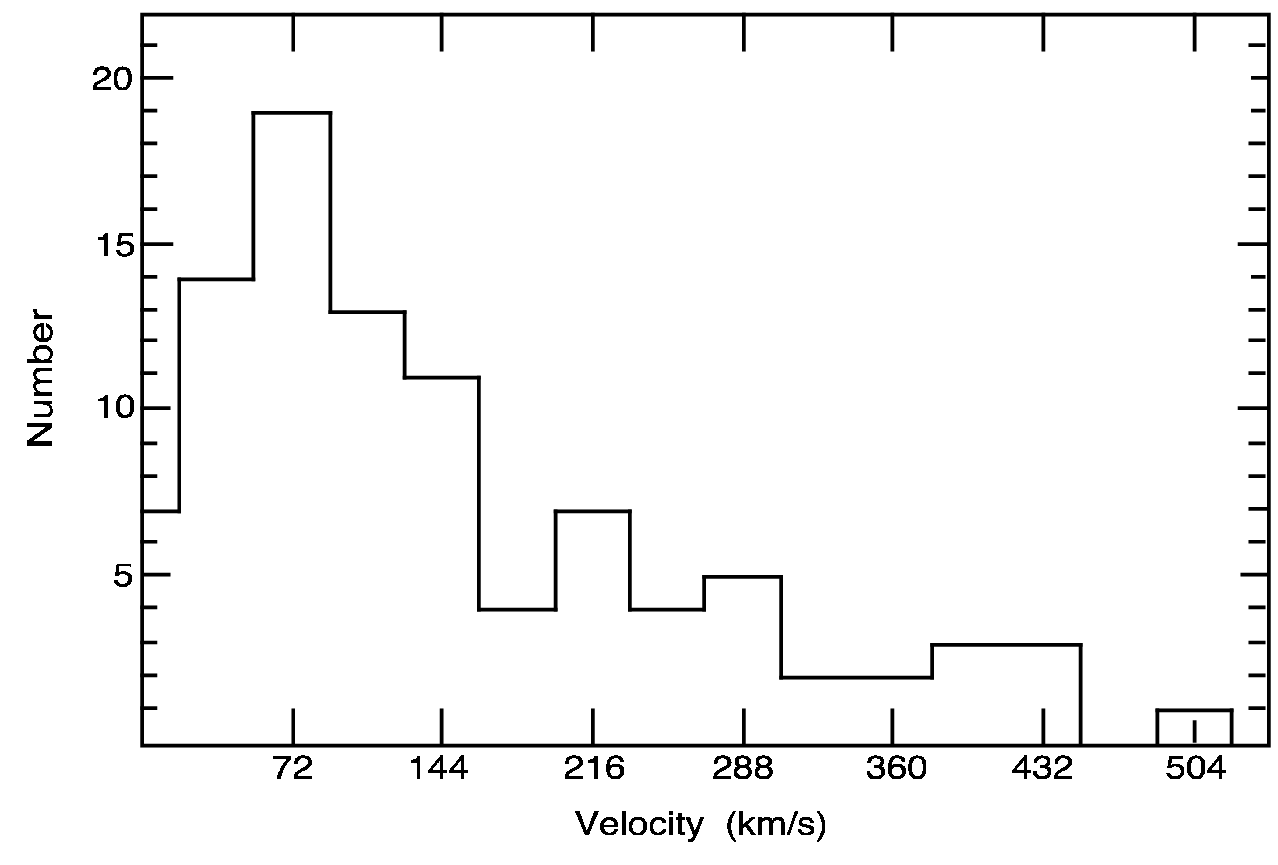}
\caption{Distribution of the absolute values of stellar velocity components near the galactic center from the data of \cite{eckart}. The diagram mixes the Right Ascension and Declination components of velocity derived from proper motion, with radial components derived from spectral shift (for a fraction of the sample). The main probability peak lies at 72=144/2 km/s.}
\label{fig:galacticcenter}
\end{center}
\end{figure}

The study of the velocity distribution of stars near galaxy centers could reveal to be particularly interesting in the context of testing the theory. Indeed there is increasing evidence that galaxies like our's host in their center compact masses (possibly black holes) of several $10^6$ M$_{\odot}$, based on observation of Keplerian velocity-distance relations. Therefore this could allow one to put the Keplerian quantization law to the test for large values of the velocities, then for large values of the coupling constant $w$, which should ultimately reach $c$ (i.e. $\alpha_{g}=1$). We give here a first preliminary example of such a work using observations by Eckart and Genzel \cite{eckart} of the proper motions of 39 stars located between 0.04 and 0.4 pc from the Galactic center. They find that these observations supports the presence of a central mass of $2.5 \pm 0.4 \times 10^ 6$ M$_{\odot}$. Though the velocities do not yet reach high values and the error bars are large ($\approx 60$ km/s), the observed distribution of velocity components, given in Fig.~\ref{fig:galacticcenter}, is compatible with a $w/n$ quantization matching with the 144 km/s sequence.

\subsection{Planetary Nebulae}
Planetary nebulae, despite their misleading names, are a general stage of evolution of low mass stars. They result from the ionization by the radiation of the central star of previously ejected outer envelopes.

The standard theory to explain the formation of planetary nebulae (PNe) is 
the interacting stellar wind (ISW) model. The simple cases of spherical and 
elliptical PNe are easily understood by this model, but many far more complicated morphologies have now been discovered, in particular thanks to high resolution Space Telescope images. Despite many attempts using numerical simulations made with a large number of initial conditions and accounting for several physical effects (magnetic fields, companion star perturbation, collimated outflows, ...), no coherent and simple understanding of these shapes has yet emerged.

Our approach to this problem is a scale-relativistic generalization of the ISW model \cite{darocha00}. Namely, we account for the chaotic motion of the ejected material and we simply replace the standard equation of dynamics used in the model by the generalized one (written in terms of the covariant derivative). It becomes a Schr\"odinger equation having well-defined angular solutions $\psi(\theta,\phi)$. Their squared modulus $P=|\psi^{2}|$ is identified with a probability distribution that presents maxima and minima (see Figure below): this means that we automatically find, by this method, that the star has a tendency to eject matter along certain angle values that are far more probable than others. Therefore we are able to predict the existence of discretized possible morphologies, in correspondence with quantized values of the square angular momentum $L^2$ and of its projection $L_{z}$.

This approach is not contradictory with the standard ones: on the contrary, for each PN with a given morphology, it will remain needed to understand why specific values of $L^2$ and $L_{z}$ is achieved. But the new point here is that the variability of shapes and their non-spherical symmetry (while the field is spherically symmetric) may now be simply understood in terms of states of a fundamental conservative quantity, the angular momentum.

Moreover, observations indicate that the propagation velocity is nearly constant (this result is already used in 
numerous simulations \cite{corra}, \cite{dwar}). This means that the PNe shells have an effective free Galilean motion. Therefore we can apply to the problem of their structuration the theoretical approach developed in the third section for a constant potential. 

\subsubsection{Elementary morphologies}
The global shaping is now understood as a consequence of the geometry of geodesics, whose distribution is described by the generalized Schr\"odinger equation (\ref{freeSE}). Its solutions, $\psi_{nlm}(r,\theta,\phi)=R_{nl}(r).Y_{l}^{m}(\theta,\phi)$ have two separable parts: the 
radial part (\ref{freeradial}) gives us information about matter density along the 
structure and the angular part imposes global shape specificity.  
The visualization of the 3D probability distribution of spherical 
harmonics gives a first result on the angular and matter density 
repartition for PNe (see figure below):
\begin{figure}[!ht]
\begin{center}
\includegraphics[width=10cm]{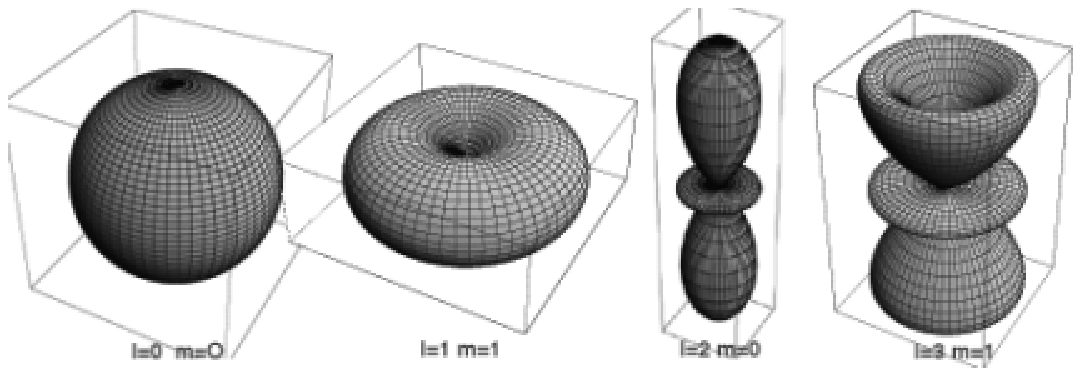}
\end{center}
\end{figure}

The model can also take into account perturbative terms, such as second order terms in the velocity power series expansion, external influences, the magnetic contribution in 
the ejection process, etc... Many hydrodynamic simulations neglect the 
magnetic force \cite{dwar,frank}, though PNe are expected to have strong magnetic fields such as the red giant stars and the white dwarfs \cite{Gseg}. The easiest description consists of
introducing a poloidal field. With this particular geometry, the flow 
of ionized particles (because of the UV star radiation) should be 
deviated in the same direction, i.e. toward the axis of symmetry.  Moreover, for particular $(l,m)$ values, bipolar structures naturally emerge. In these singular objects, a self-gravitating force appears. 
This force acts like the magnetic field and induces a constriction along the axis of symmetry of the PNe. 
So, the angular distribution has to be corrected to account for the 
perturbations introduced. 

\subsubsection{Morphological results}
We present a synthesis of the different shapes allowed by our 
model. Three categories resume all the possibilities: \\
$\bullet$ Spherical and elliptic: The basic spherical shape is 
obtained for the specific value of $(l=m=0)$. For the restriction of 
($[m=\pm l]\,\forall\,l\ge 1$), PNe will evolve to an equatorial disc. 
The tilting on the line of sight of this disc will induce a 
spherical or an elliptical PN.  
\begin{figure}[!ht]
\begin{center}
\includegraphics[width=6cm]{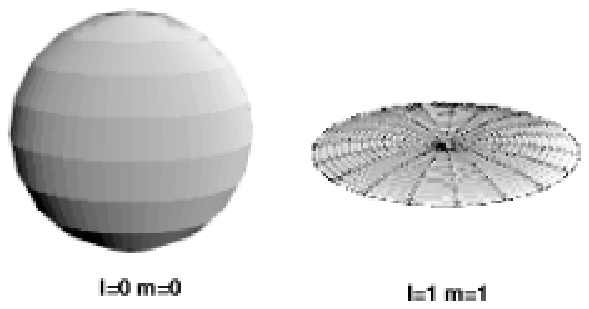}
\end{center}
\end{figure} \\
$\bullet$ Bipolar ejection: The general process upon which our whole description is based is an ejection process. Therefore, it is not surprising to find solutions 
describing bipolar jets ejection. For ($[m=0]\,\forall\,l\ge 1$), the density 
distribution is concentrated on the axis of the objects.
\begin{figure}[!ht]
\begin{center}
\includegraphics[width=10cm]{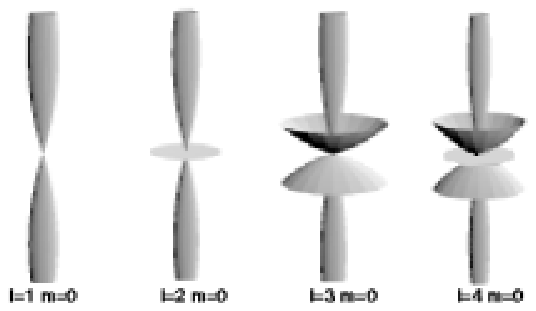}
\end{center}
\end{figure} \\
$\bullet$ Bipolar shell: All the other solutions give bipolar shell 
structures. The empirical relation $(l-m)+1$ constrains the number of internal structures. For example, ($l=6,\,m=2$) gives $5$ structures (one disc and four shells). 
\begin{figure}[!ht]
\begin{center}
\includegraphics[width=12cm]{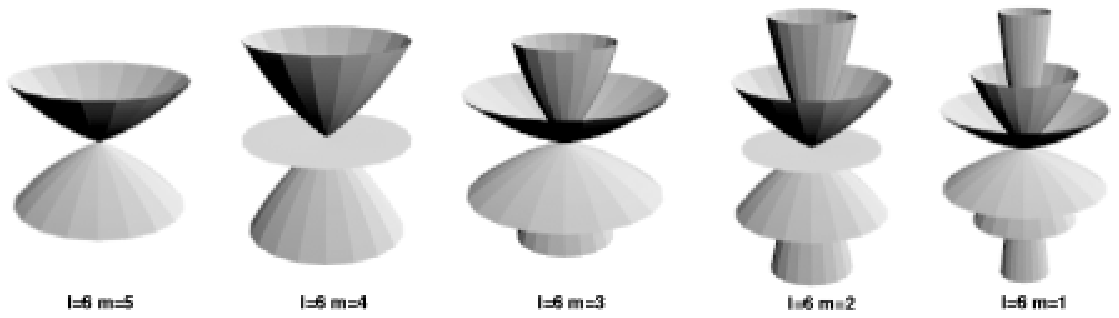}
\end{center}
\end{figure}
This brief presentation allows one to classify all the elementary shapes and gives a method for constraining the structure with the couple $(l,m)$.

\subsubsection{Comparison and discussion}
\begin{figure}[!ht]
\includegraphics[width=15cm]{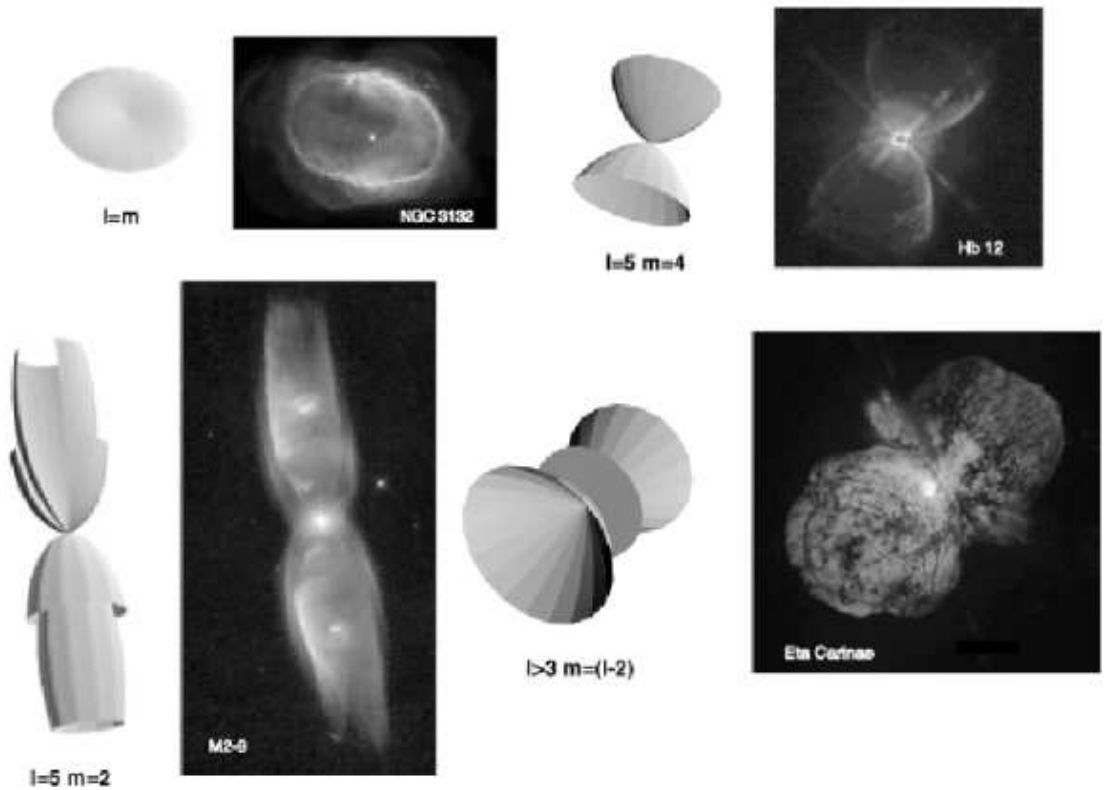}
\caption{Direct comparison between predicted quantized shapes and typical observed Planetary Nebulae, plus the $\eta$ Carinae.nebula (STScI images, adapted from http://ad.usno.navy.mil/pne/caption.html).}
\label{morpho}
\end{figure}

\begin{figure}[!ht]
\includegraphics[width=15cm]{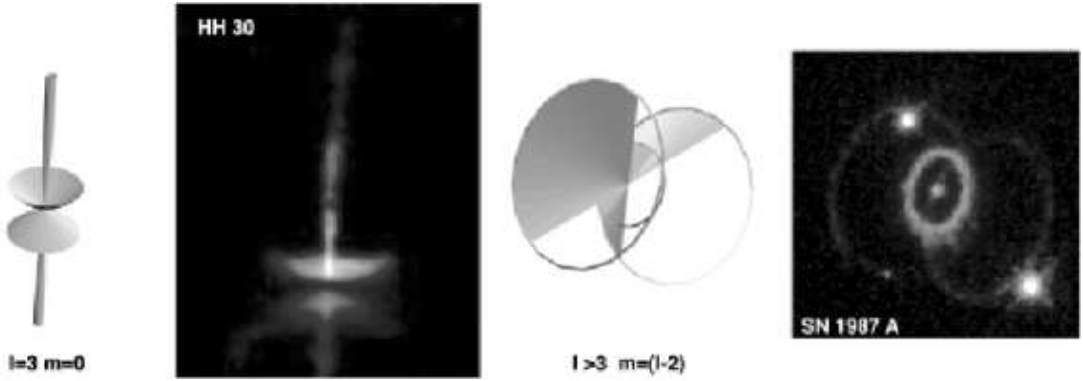}
\caption{Generalization of the theoretical description of ejection process. The left hand side represents a young star ejection/accretion state, and the right hand side, a supernova explosion (adapted from STScI images).}
\label{morpho2}
\end{figure}

The following four examples (Fig.~\ref{morpho}) show the direct comparison between structures observed and predicted quantized shapes built with the Schr\"odinger model: we can see that many exotic shapes are naturally obtained in this framework.
Moreover, the general solutions could be used in many ejection/scattering cases other than PNe. In Fig.~\ref{morpho2}, we compare the shapes of young star ejection states or SuperNovae explosions with quantized solutions (also valid for inward motion).
To conclude this section, the Schr\"odinger approach brings a new conception in the general ejection process in astrophysics. Our PNe shaping model treats in the same way all the elementary shapes and is in accordance with many singular observations.

\subsection{ Extragalactic structures}

\subsubsection{Rotation curves of spiral galaxies}
The flat rotation curves observed in the outer regions of spiral galaxies is one of the main dynamical effects which demonstrates the so-called ``missing mass" problem. Indeed, farther than the visible radius of galaxies, a very small quantity of matter is detected by all the possible methods used (images at all wavelengths, gravitational lensing, 21 cm radio observations, etc...). Therefore one expects the potential to be a Kepler potential $\phi \propto 1/ r$ beyond this radius, and as a consequence one expects the velocity to decrease as $v \propto r^{-1/2}$, while all available observations show that the velocity remains nearly constant up to large distances for all spiral galaxies. In order to explain this effect and other similar effects at all extragalactic scales (in particular in clusters of galaxies, as first discovered by Zwicky in the years 1930), one usually makes the hypothesis of the existence of huge quantities of missing, ``dark" matter that would be the dominant mass density constituent of the Universe. However, despite decades of very active research, this missing matter continues to escape detection. Another suggestion was to modify Newton's gravity (the MOND hypothesis), but it has up to now not been possible to render such an ad-hoc hypothesis consistent with general relativity nor with observations at different scales.

Moreover, the ``dark matter" problem is deeply connected with the problem of the formation of galaxies and of large scale structures of the Universe. Indeed, in the standard approach to this problem, it would be impossible in its absence for the very small $z=1000$ fluctuations to grow toward today's structures. However, as recalled in Silk's quotation at the beginning of this contribution, even in its presence the theory of gravitational growth remains unsatisfactory.

The scale-relativity approach allows one to suggest an original solution to both problems. Indeed, the fractal geometry of a non-differentiable space-time solves the problem of formation on many scales (this is the subject of this whole contribution). It also implies the appearance of a new scalar potential (Eq.~\ref{Q}), which manifests the fractality of space in the same way as Newton's potential manifests its curvature. We suggest that this new potential may explain the anomalous dynamical effects, without needing any missing mass. 

This will be discussed in more detail in the joint paper \cite{newcosmuniv}. Let us exemplify this result here in the case of the flat rotation curves of spiral galaxies. The formation of an isolated galaxy from a cosmological background of uniform density is obtained, in its first steps, as the fundamental level solution $n=0$ of the Schr\"odinger equation with an harmonic oscillator potential (Eq.~\ref{oscill}). Its subsequent evolution is expected to be a solution of the Hartree equation (\ref{hartree}): this will be the subject of forthcoming works. 

Once the galaxy is formed, let $r_{0}$ be its outer radius, beyond which the amount of visible matter becomes small. The potential energy at this point is given, in terms of the visible mass $M$ of the galaxy, by:
\begin{equation}
\phi_{0}=-\frac{GMm}{r_{0}}=  - m \, v_{0}^{2}. 
\end{equation}
Now the observational data tells us that the velocity in the exterior regions of the galaxy keeps the constant value $v_{0}$. From the virial theorem, we also know that the potential energy is proportional to the kinetic energy, so that it also keeps a constant value given by $\phi_{0} =-GMm / r_{0}$. Therefore $r_{0}$ is the distance at which the rotation curve begins to be flat and $v_{0}$ is the corresponding constant velocity. In the standard approach, this flat rotation curve is in contradiction with the visible matter alone, from which one would expect to observe a variable Keplerian potential energy $\phi=-GMm/r$. This means that one observes an additional potential energy given by:
\begin{equation}
\label{qobs}
Q_{obs}=-\frac{GMm}{r_{0}} \left(  1-  \frac{r_{0}}{r} \right)  . 
\end{equation}
Now the regions exterior to the galaxy are described, in the scale-relativity approach, by a Schr\"odinger equation with a Kepler potential energy $\phi=-GMm/r$, where $M$ is still the sole visible mass, since we assume here no dark matter. The radial solution for the fundamental level is given by:
\begin{equation}
\sqrt{P}=2 \; e^{-r/r_{B}}, 
\end{equation}
where $r_{B}=GM/w_{0}^{2}$ is the macroscopic Bohr radius of the galaxy.

\begin{figure}[!ht]
\begin{center}
\includegraphics[width=8cm]{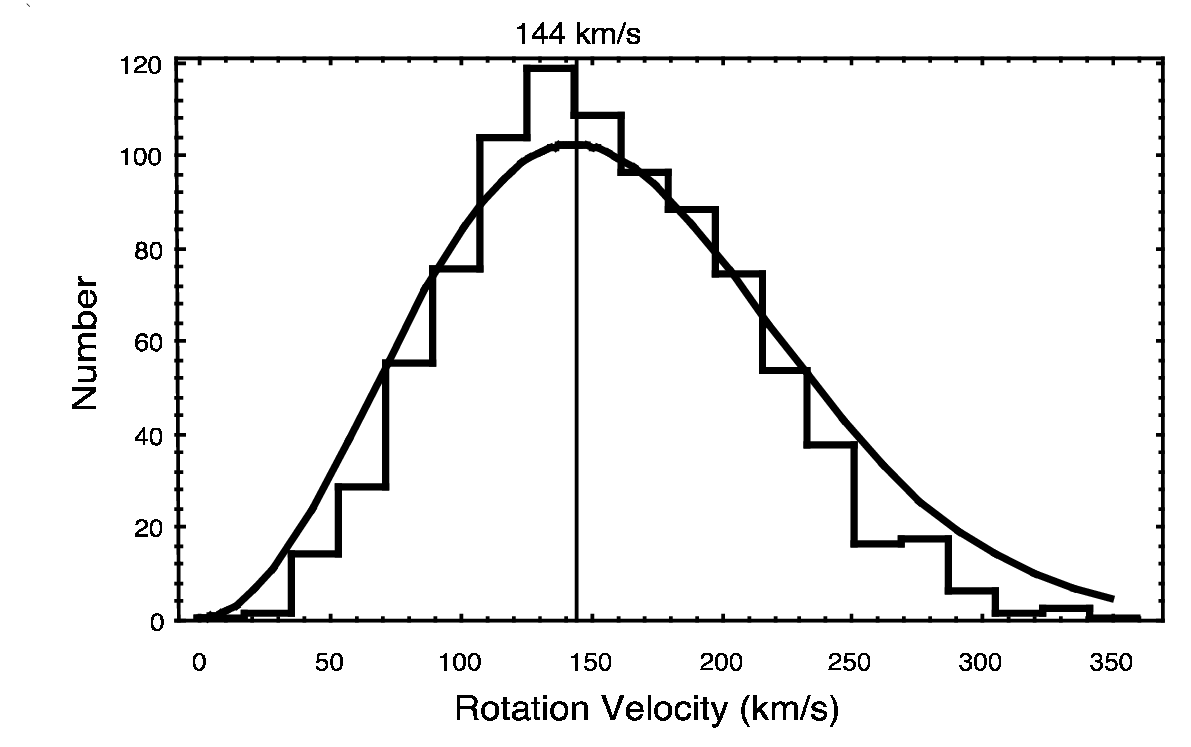}
\caption{Distribution of the outermost observed velocities in spiral galaxies (flat rotation curves) from the catalog of rotation curves for 967 spiral galaxies by Persic and Salucci \cite{persic}. The fitted continuous curve is proportional to $v^2 \,  \exp\left(-(v/144)^2\right)$.}
\label{fig:rotationcurve}
\end{center}
\end{figure}

It is now easy to compute the theoretically predicted form of the new potential (Eq.~\ref{Q}), knowing that ${\cal D}=GM/2w_{0}$:
\begin{equation}
Q_{pred}=-2 m \, {\cal D}^{2} \; \frac{\Delta \sqrt{P}}{\sqrt{P}}= -\frac{GMm}{2r_{B}} \left(  1-  \frac{2r_{B}}{r} \right) =-\frac{1}{2} \, w_{0}^{2}\,  \left(  1-  \frac{2r_{B}}{r} \right). 
\end{equation}
We therefore obtain, without any added hypothesis, the observed form (Eq.~\ref{qobs}) of the new potential term. Moreover the visible radius and the Bohr radius are now related, since the identification of the observed and predicted expressions yield: $r_{0}=2 \, r_{B}$. The constant velocity $v_{0}$ of the flat rotation curve is also linked to the fundamental gravitational constant $w_{0}$ by the relation $w_{0}= \sqrt{2}\, v_{0}$. This prediction is consistent with an analysis of the observed velocity distribution of spiral galaxies from the Persic-Salucci catalog \cite{persic}, as shown in Fig.~\ref{fig:rotationcurve} \cite{pasty}. The peak velocity is $142 \pm 2$ km/s, while the average velocity is found to be $156 \pm 2$ km/s, so that $\sqrt{2} <v>=220 \pm 3$ km/s: this value, which we shall also obtain in the study of clusters of galaxies (see below) is within 1$\sigma$ of $3/2 \times 144.7$ km/s.

\subsubsection{Compact groups}
As remarked by Hickson \cite{hicksonCAT}, ``compact groups of galaxies provide the best environment for galaxy interactions to occur, so it is natural to study these systems in order to better understand the interaction process and
its effects". ``But the existence of such systems is a puzzle, as they are unstable to gravitational
interactions and mergers". ``A way out of this impasse is to assume that new compact groups are constantly forming". Finally Hickson reaches the conclusion, with now many other specialists, that ``in compact groups such as those illustrated [in the Atlas of Compact Groups of Galaxies], we may be actually observing the process of galaxy formation".

\begin{figure}[!ht]
\begin{center}
\resizebox{8cm}{!}{\includegraphics{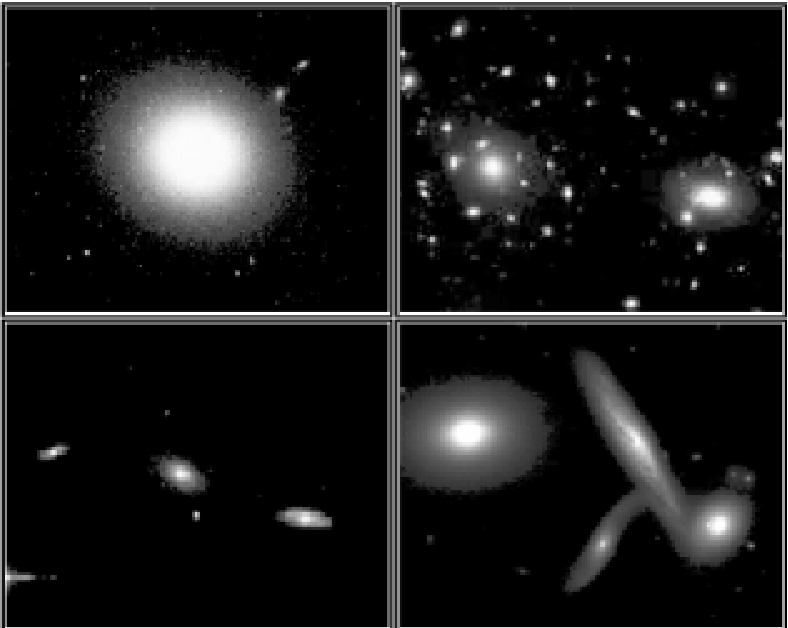}}
\caption{\footnotesize{Galaxy grouping: (up left) the giant elliptical galaxy M87 in the center of the Virgo cluster (photo AAT, D. Malin); (up right) the center of the Coma cluster; (down) two examples of typical compact groups of galaxies : the triplet HCG 14 \cite{hicksonWEB} and a trapeze structure in the quintet HCG 40 \cite{subaru}.}}
\label{fig:compactgroup}
\end{center}
\end{figure}
It is therefore remarkable in this context that the kind of morphologies displayed in a systematic way by these compact groups is precisely chain and quadrilateral structures quite similar to those encountered in star forming regions (see Fig.~\ref{fig:compactgroup}). The basic solutions obtained from the gravitational Schr\"odinger equation for formation from a background of constant matter density (i.e. harmonic oscillator potential) once again explain these morphologies in a very simple way.

\subsubsection{Galaxy pairs}
In the same way as binary stars in our Galaxy, binary galaxies are very common structures in the Universe. For example our own Local Group of galaxies (which deserves a special study at the end of this paper) is organized around the pair of giant spirals (Milky Way Galaxy / Andromeda nebula). This is easily understood in our framework, since double structures are the lowest energy solution (beyond the fundamental level that represents an isolated object) of the gravitational Schr\"odinger equation with a harmonic oscillator potential.

Moreover, binary galaxies are one of the first extragalactic systems for which a redshift quantization effect in terms of 144km/s and its submultiples has been discovered by Tifft \cite{tifft77}. However, this effect has been interpreted by Tifft and other authors as an ``anomalous redshift" of non-Doppler origin, whose existence would therefore question the whole foundation of cosmology. But in such an interpretation, there should be a fundamental difference of behavior between motion deduced from extragalactic redshifts and the motion of planets. The discovery \cite{xtrasol,revueFST,agnese}, motivated by the scale-relativity predictions, that the planets of our own Solar System and of extra-solar planetary systems do have velocities involved in the same sequence $v_{n}=(144/n)$ km/s (namely, the velocities of Mercury, Venus, the Earth and Mars are respectively $\approx 48$, 36, 29 and 24 km/s) has definitively excluded the anomalous redshift interpretation.

On the contrary, not only the extragalactic redshifts are therefore confirmed to be of Doppler and cosmological origin, but the preferential velocity values take meaning here as a mere manifestation of the formation of structures. In the same way as there are well-established structures in the position space (galaxies, groups, clusters, large scale structures), the velocity probability peaks are simply the manifestation of structuration in the velocity space.

\begin{figure}[!ht]
\begin{center}
\includegraphics[width=8cm]{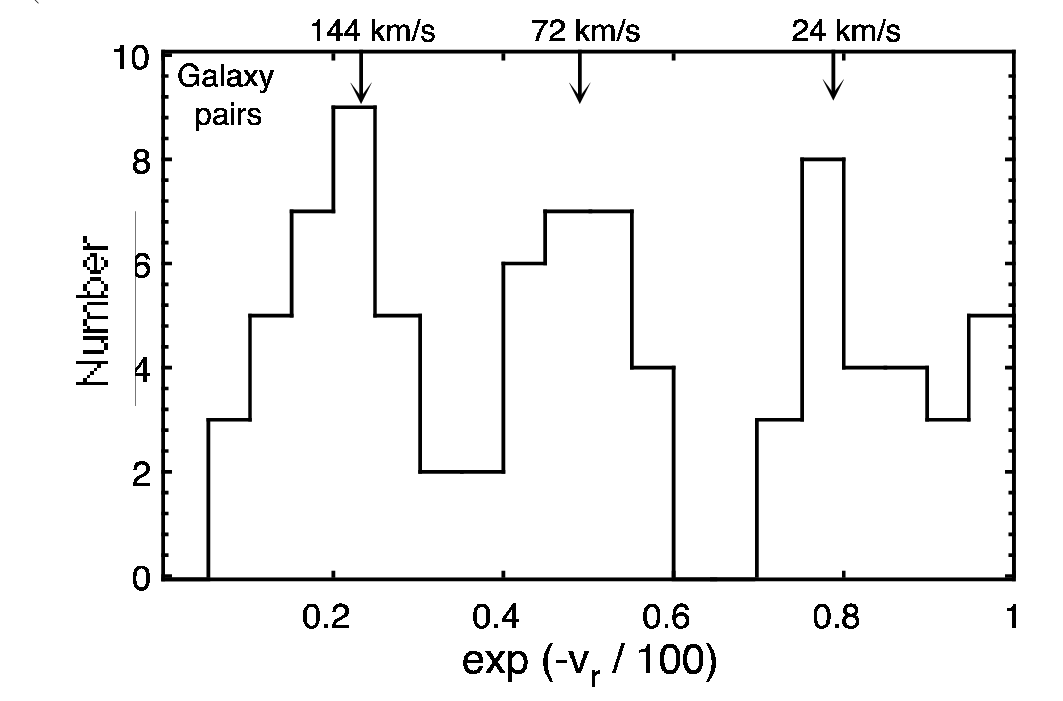}
\caption[]{Deprojection of the intervelocity distribution of galaxy pairs \cite{bingal} from the Schneider-Salpeter catalog with precision redshifts \cite{schneider92}. The two main probability peaks are found to lie at 144 and 72 km/s.}
\label{fig:galaxypairTN}
\end{center}
\end{figure}

In a recent work, several methods of deprojection of the intervelocity (only the radial component is observed) and of the interdistance of binary galaxies (only the two transverse components are observed) have been developed \cite{bingal}. The result confirms the existence of probability peaks in the velocity space (see Fig.~\ref{fig:galaxypairTN}) and in the position space (more precisely, in the interdistance to mass ratio distribution), in agreement with the scale-relativity prediction for the Kepler potential of the pairs in reduced coordinates. These peaks are correlated through Kepler's third law, which is a final demonstration of the Doppler origin of the redshift differences in galaxy pairs. 

\subsubsection{Clusters of galaxies}
\begin{figure}[!ht]
\begin{center}
\includegraphics[width=16cm]{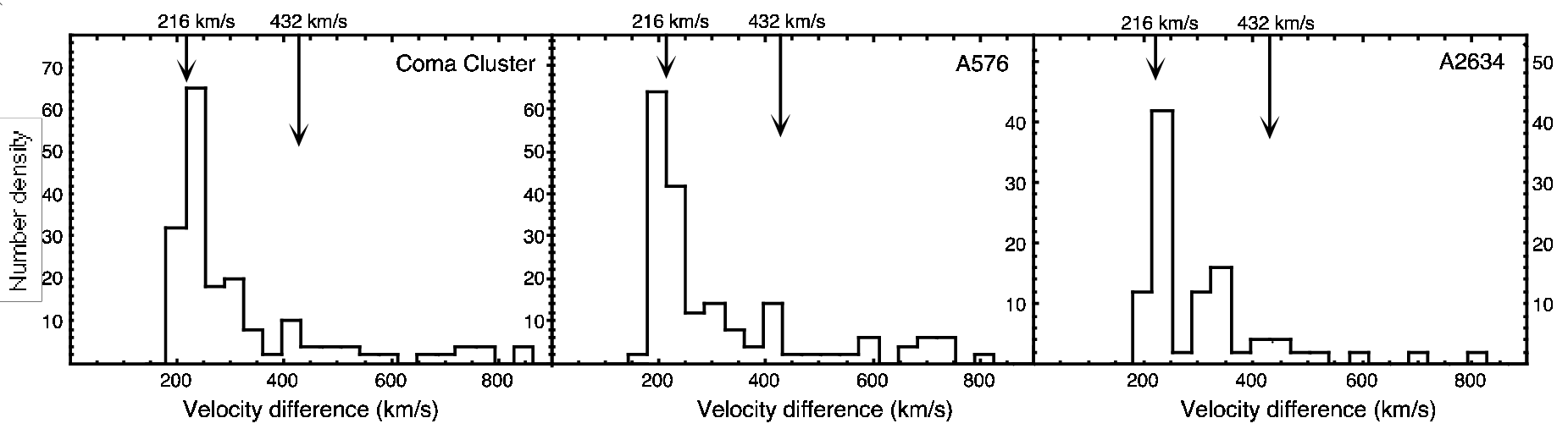}
\caption[]{Histogram of the modulus of the discrete Fourier transform of radial velocities of galaxies in three clusters, Coma (230 redshifts), A576 (221 redshifts) and A2634 (125 redshifts). A significant peak around $\approx 216 =3 \times 72$ km/s is found in every cases.}
\label{fig:cluster}
\end{center}
\end{figure}

The Coma cluster of galaxies is the first system in which probability peaks in the redshift distribution have been discovered \cite{Tifft76}, in units of $\approx 216$ km/s. Recall that we have also identified such a value in our own Solar System, since it corresponds to the Kepler velocity of infra red dust at 4.09 solar radius (see Fig.~\ref{fig:IRdust}). We give in Fig.~\ref{fig:cluster} the result of an analysis using a discrete Fourier transform of more recent accurate redshifts for three clusters, Coma, A576 and A2634. We recover the Tifft result for Coma and confirm it with A 576 and A2634.

A detailed application of the scale-relativity theory to these structures is still to be developed. Indeed, we deal here with a self-gravitating system that come under the Schr\"odinger-Poisson coupled equations, or, equivalently, under the Hartree equation (which is of degree 4). Moreover, a global description in terms of a quantum-like statistical physics is also needed in this case. This will be the subject of a forthcoming work.

\subsubsection{Local Supercluster}
\begin{figure}[!ht]
\begin{center}
\includegraphics[width=6cm]{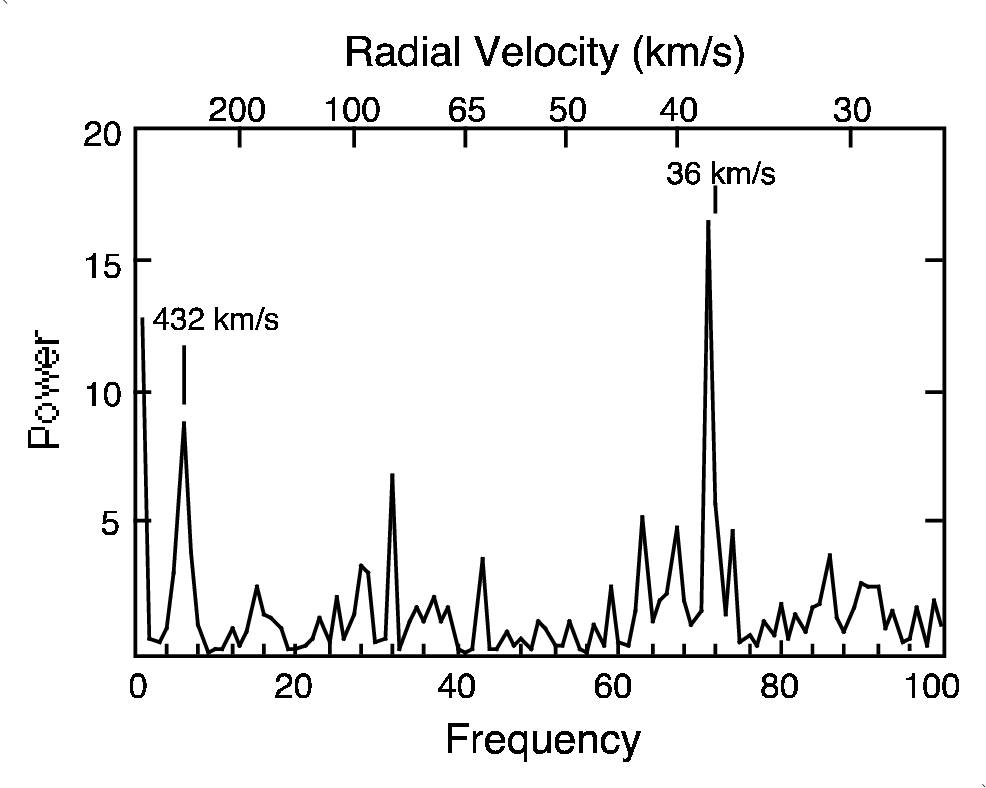}
\caption[]{Power spectrum of the radial velocities of the 97 galaxies in the Guthrie-Napier sample with accurately determined redshifts, corrected for the optimum solar vector (219 km/s, 96 degree, $-11$ degree). The total range of velocities is 2665 km/s. Significant peaks are obtained for $v \approx 36$ km/s and $v \approx 432$ km/s.}
\label{fig:GuthrieNapier}
\end{center}
\end{figure}

 The existence of a preferential value of $\approx 36=144/4$ km/s for the intervelocity of galaxies at the scale of the local supercluster was first suggested by Tifft and Cocke \cite{TC84}. Croasdale \cite{croasdale}, then Guthrie and Napier \cite{GN91} found some support for this claim using spirals with accurately measured redshifts up to $\approx 1000$ km/s. In a more recent work \cite{GN96}, they have confirmed the effect with galaxies reaching $\approx 2600$ km/s. We have performed a power spectrum analysis on their database (see Fig~\ref{fig:GuthrieNapier}): we indeed find a highly significant peak corresponding to a preferential intervelocity of 37.5 km/s, and also a marginally significant one at 432 km/s.

\begin{figure}[!ht]
\begin{center}
\includegraphics[width=8cm]{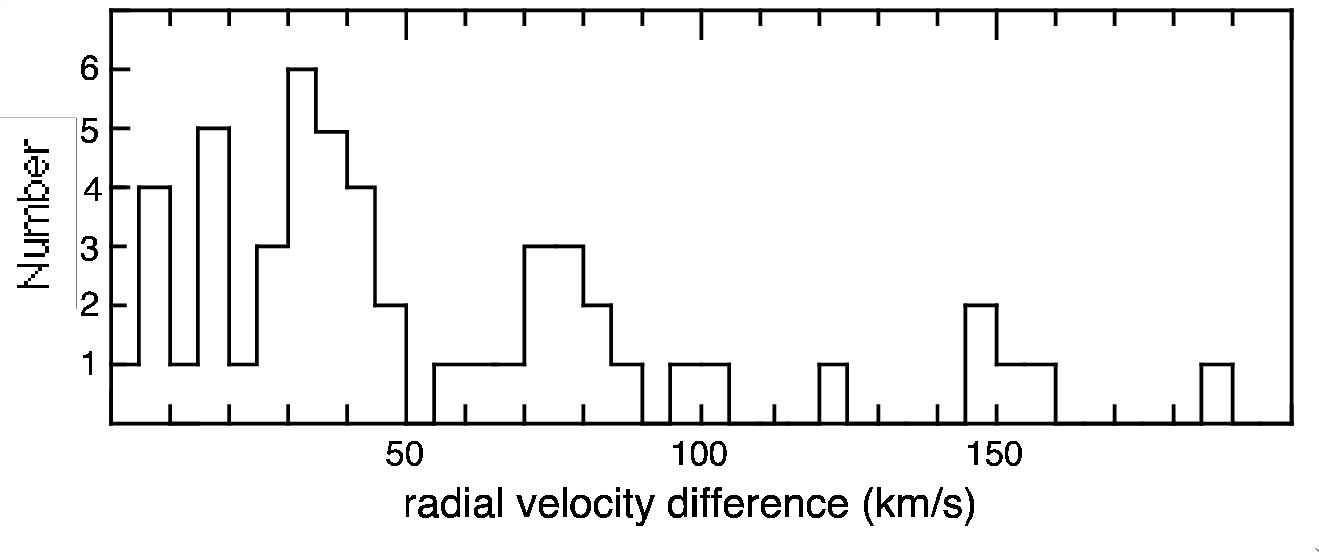}
\caption[]{Distribution of closest intervelocities of galaxies which are members of groups in the Guthrie-Napier accurate redshift sample.}
\label{fig:supercluster}
\end{center}
\end{figure}

Moreover, Guthrie and Napier remark that the phenomenon appears strongest for the galaxies linked by group membership. We confirm this result by a specific analysis of the group galaxies (Fig~\ref{fig:supercluster}), which show peaks of their intervelocity distribution at $\approx144$, 72, 36 km/s and possibly sub-levels (while $108=3 \times 36$ km/s is absent), as expected for a local Kepler potential for which $v_{n}=w_{0}/n$. It is also supported by the study by Jacob \cite{jacob} and Lefranc \cite{lefranc} of other independent samples of galaxies at the scale of the local supercluster, which have provided significant peaks at 48 and 36 km/s.

All these properties can be readily explained in the scale-relativity framework. Indeed, the potential to be inserted in the gravitational Schr\"odinger equation for describing the structuration of these objects is, to the first order approximation, a combination of a 3-D harmonic oscillator potential describing the density background at the scale of the local supercluster and of local Kepler potentials in the grouping zones. The solution $n=1$ of the harmonic oscillator leads one precisely to expect the existence of a preferential intervelocity depending on the density and on the gravitational coupling constant. Moreover, since these structures are understood as the equivalent in velocity space of the well-known structures in position space (galaxies, groups, clusters, etc...), we also expect their strengthening in groups and the local replacement of a linear quantization law by an inverse law.

\subsubsection{Very large scale structures}

The morphology of superclusters of galaxies is, like stellar associations, ensembles of star clusters and groups of galaxies, characterized by the existence of pairs, chains, etc..., as expected from the solution of the Schr\"odinger equation in the cosmological potential (i.e., harmonic oscillator in de Sitter coordinates) produced by a uniform background matter density.

Redshift structures of the kind that scale-relativity is expected to predict, namely, probability peaks for some specific values of redshift differences, have also been detected on very large scales. Broadhurst et al. \cite{broadhurst} have detected a pseudoperiodicity in units of 12800 km/s in the distribution of galaxies in two opposite cones directed toward the North and South Galactic poles. This effect has since that time been confirmed in several redshift samples of galaxy and clusters of galaxies (see e.g. \cite{einasto}).

At even larger scales, the existence of preferred redshifts for quasars has been pointed out by Burbidge and Burbidge \cite{burbidge} and subsequently confirmed by many authors. One finds peaks for redshift differences such that $\Delta\ln(1+z)=0.206$ \cite{karlsson}.

The detailed understanding of such an effect needs an application of the scale-relativity approach to the cosmological realm, since the corresponding length-scale of the Broadhurst et al. effect is about 180 Mpc (using the recently measured precise Hubble constant $H_{0}=70 \pm 7$ km.s$^{-1}$.Mpc$^{-1}$), while the gigaparsec is almost reached with the QSO effect. Such a study is too long to be undertaken here. Let us only briefly note that we expect the gravitational coupling constants $\alpha_{g}=w/c$ at different scales to be interrelated, and, ultimately, we expect the $w$'s to be related to the maximal velocity $c$ (i.e., $\alpha_{g}=1$). 

The observed values of the large scale and small scale constants are consistent with such a matching. Indeed, the inverse coupling constant corresponding to $w_{0}=144.82 \pm 0.01$ km/s (see \cite{xtrasol2} for a suggested theoretical prediction of this value) is $\alpha_{g}^{-1}=2070.10 \pm 0.15$. Since $2070=2 \times 3^2 \times 5 \times 23$, it is remarkable that $c/23=13034$ km/s, which is compatible with the Broadhurst et al. periodicity, while the quasar periodicity is close to $c/5=0.2$. Moreover, various multiples of $w_0$ (by factors of 2, 3 and 9) which have been observed in several systems also belong to the same sequence.

\subsection{Local Group of galaxies}

Let us conclude this contribution by a more detailed analysis of our Local Group of galaxies. It is a particularly interesting system as concerns the application of the gravitational Schr\"odinger equation, since it is essentially made of two giant spiral galaxies (our Milky Way galaxy and M31) surrounded by their companions. 

The analogy of its structure (shared with several other loose groups of spirals) with old expanding stellar associations has been pointed out long ago by de Vaucouleurs \cite{vaucouleurs}. As we have seen in previous sections, the theory of scale relativity allows one to understand the gravitational formations of such binary systems (and of multiple ones) in terms of the first excited state ($n=1$) which is solution of the gravitational Schr\"odinger equation for a constant background density (harmonic oscillator potential). 

The use of the scale-relativity approach in this case is also supported by the investigation of the motion of these galaxies in numerical simulations (see e.g. \cite{valtonen}) that has demonstrated the chaotic and violent past and future history of the Local Group. Moreover the loose character of this group implies a velocity field which is locally dominated by the gravitation of the two giant spirals, but which is expected to rejoin the Hubble expansion field in its outer regions. The expected quantization law is therefore rather complicated in this case, since it should correspond to a Kepler potential near M31 and MW, then to a two-body potential in an intermediate region, and finally to a harmonic oscillator potential at the scale of the local supercluster. A possible redshift quantization in the Local Group in units of 72 km/s has already been detected \cite{arp86}.

Observations show that there is a net age difference between both dominant galaxies and the rest of the dwarf galaxies in our Local Group \cite{mateo}. One can also assert that the gas is isotropic in each subgroup and is subjected to the simple Keplerian potential of the dominant galaxies. Thus, we can use the Keplerian solutions developed in the third section. All the solutions should be constrained in order to agree with the initial system (spherical symmetry and isotropic subgroups). The isotropic information is contained in the angular part of the equation (\cite{bah}). Spherical harmonics, $Y _ { l } ^ { \hat{m} } (\theta, \phi) $, reveal an isotropic arrangement only for $l=0$ and $\hat{m}=0$. Then the mean distance to the gravitational center is 
given by the formula ${ < r >} _ { n } = (3GM / 2 w_{0}^{2}) n ^ {2}$ (cf. Keplerian problem section).
This equation assumes a particular quantization law (in $n^{2} $) for the galactic 
distances with regard to the center of the dominant galaxies. The two dominant galaxies will therefore infer two different laws in two different domains (in this first Keplerian stage). The 
constant $w_{0}$ can not be fixed beforehand : a first step then consists of 
considering the main constant at $144$ km.s$^{-1}$ and the closest values ($288$ km.s$^{-1}$, $72$ km.s$^{-1}$). 
The mass $M$ is the visible mass of the Milky Way. From the data 
of Bahcall \cite{bah}, we take a representative mass of 
$7.2\times10 ^ {10} M_{\odot} $. For M31, from the data used by Kent et al. \cite {kent}, it seems 
reasonable to take a visible mass value close to $13.2\times10 ^ {10} M 
_ {\odot} . $ \\
For these two systems, we can estimate the laws of the Keplerian model allowed by the scale-relativity approach. Various values of $w _ {0} $ are considered:

\begin{table}[!ht]
\begin{center}
\begin{tabular}{|c|c|c|c|}
\hline
$w_{0}$&72 km.s$^{-1}$&144 km.s$^{-1}$&288 km.s$^{-1}$\\
\hline
$\frac{3}{2} \frac{GM_{MW}}{w_{0}^{2}}$&89.2  kpc& 22.3  kpc&5.57 
$$ kpc\\
\hline
$\frac{3}{2} \frac{GM_{M31}}{w_{0}^{2}}$ & 164  kpc& 40.1  kpc & 
10.25  kpc \\
\hline
\end{tabular}
\end{center}
\caption{Theoretical characteristic distances for the Milky Way and M31 systems in function of $w_{0}$.}
\end{table}

Furthermore, it is necessary to treat the case of the remote galaxies and of 
the NGC 3109 galaxies subgroup. From the theoretical point of view, this is an interesting problem since, assuming a global coherence of the double system in its outer regions described in terms of a global ${\cal D}$ value, it shares some common features with the Schr\"odinger equation written for molecules like $H_ {2} ^ {+} $ (namely, a test particle subjected to two attractive centers). One finds that the wave function, solution of such a problem, is $ \psi= a_1 \,  \psi_1(r _ {1}) + a_2\, \psi_2 (r _ {2}) $, where $ \psi_1$ and $ \psi_2$ are solutions of the Sch\"odinger equations written using this global ${\cal D}$ for each individual Kepler potential. The global solution can be subsequently matched with the local solutions. Unfortunately, the 
case of the Local Group (asymmetrical gravitational double system) is more 
complicated because $ {\cal D} _ { MW } \ne {\cal D} _ {M31} $ and $ 
{\psi} _ {MW} \ne \psi_ {M31} $. Nevertheless, the molecular solution 
has the advantage to supply a simple presence probability, revealing 
the interference of the individual solutions: ${\cal P}= 
a_1^2{\cal P}_1^{2}+a_2^2 {\cal P}_2^2+ 
2 a_1 a_2 \sqrt{{\cal P}_1 {\cal P}_2} \cos(\Delta \theta)$. Even if this solution can not be used as such in the more complex macroscopic gravitational case, it will be interesting to look at the configuration of 
the remote galaxies with regard to the laws that apply around M31 and Milky Way.

\subsubsection {Structures in position space}
The precision on the distances and the radial velocities in Mateo's synthesis work \cite{mateo} is sufficient for using them directly in our study.  By the knowledge of the sun vector, we 
calculate the galactocentric distances $d$. To verify the existence of a 
law of the form $d=d_{0} \times n^2$, we shall analyze the distribution of the observable $\widetilde{n}=\sqrt{d / d_{0}}$. We expect this distribution to exhibit probability peaks around integers values ($n$). 

\paragraph{Milky Way subgroup}
The study of the data about the distances of the close-by galaxies reveals a minimum for a characteristic distance $d_{0}=5.50$ kpc. This result is in very good agreement with the theoretical prediction of $5.57$ kpc for $w_{0}=288$ km.s$^{-1}$. We therefore compute the   values of $\widetilde{n}=\sqrt{d / 5.57}$, then the differences between them and the nearest integer ($\delta n =\widetilde{n}-n$). In the standard framework one expects these differences to be uniformly distributed between $-0.5$ and $+0.5$, while, in the present approach, one expects them to peak around zero. 

\begin{table}[!ht]
\begin{tabular}{|c|c|c|c|c|c|c|c|c|c|c|c|c|}
\hline  & \small Sag&\small LMC&\small SMC&\small Ursa&\small 
Sculp&\small 
Drac&\small Sext&\small Cari& \small Forn&\small Leo II&\small 
Leo I&\small 
Phoe \\
\hline
\small d&\small 16&\small 48&\small 55&\small 68&\small 79&\small 
82&\small 89&\small 103&\small 140&\small 207&\small 254&\small 445\\
\hline
$ \widetilde{n}=\sqrt{\frac{d}{5.50}}$&\small 1.70&\small 
2.95&\small 
3.16&\small 3.51&\small 3.79&\small 3.86&\small 4.02&\small 4.32&5.02& 
\small 6.13&\small 6.79&\small 8.99\\
\hline
$\delta n=\widetilde{n}-n$&\small -0.30&\small -0.05&\small 
0.16&\small -0.49&\small -0.21&\small -0.14&\small 0.02&\small 
0.32&\small 
0.02& \small 0.13&\small -0.21&\small -0.01\\
\hline
\end{tabular}
\begin{center}
\begin{tabular}{cc}
\includegraphics[width=4cm]{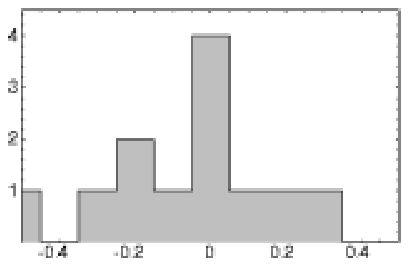} &
\includegraphics[width=4cm]{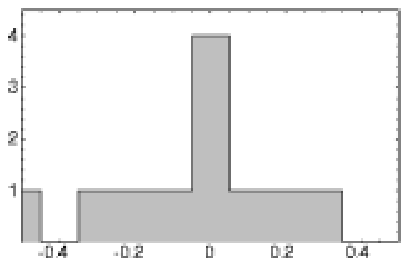} \\

\end{tabular}
\end{center} 
\caption{Histograms of the difference between the variable $\widetilde{n}=\sqrt{d / 5.57 \, kpc}$ and its nearest integer, for the Milky Way companion galaxies. The right-hand side represents the MW subgroup, Leo I being excluded due to the large uncertainty on its distance.}
\end{table}

The result, shown in the tables and histograms, clearly favors the gravitational Schr\"odinger approach, despite the small number of objects. 

\paragraph{Andromeda subgroup}

The distance distribution of M31 companion galaxies is also in good agreement with a quantized $n ^ {2}$ distribution, for a value of $d_{0}=10.72$ kpc. Once again, the Keplerian model developed for $w_{ 0 }=288$ km/s (that gives an expected value $d_{0}=10.25$ kpc) is close to the observed law in $r=10.72 \, n ^ {2}$ kpc. 
\begin{table}[!ht]
\begin{tabular}{|c|c|c|c|c|c|c|c|c|c|c|c|c|}
\hline  &   M32 &  N205&  
AI&  AIII &  N147&  N185&  
M33&  IC10&   AII&  LGS3&  
IC1613&  EGB\\
\hline
  d&  35&    45&    57& 
   66&    106&    172&    
219&    261&    270&    275& 
   500&  685\\
\hline
$  \widetilde{n}$&  1.80&    2.04& 
   2.30 &   2.48&    3.14& 
   4.00&    4.52&    4.93& 
   5.01&    5.06&   6.82&   
7.99\\
\hline
$  \delta n$&  -0.20&    0.04& 
   0.30&    0.48&    0.14& 
   0.00&    -0.48&    -0.07& 
  0.01&   0.06&    -0.18&  -
0.01\\
\hline
\end{tabular}
\begin{center}
\begin{tabular}{cc}
\includegraphics[width=4cm]{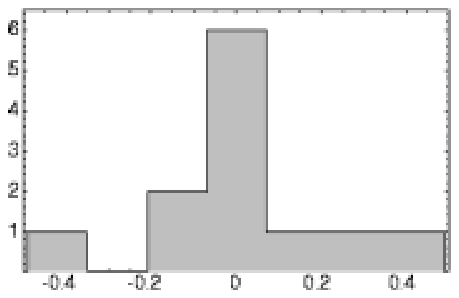} &
\includegraphics[width=4cm]{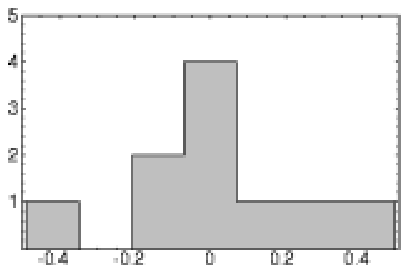} \\

\end{tabular}
\end{center}
\caption{Histograms of the difference between the variable $\widetilde{n}=\sqrt{d / 10.72 \, kpc}$ and its nearest integer, for M31 companion galaxies. The right-hand side represents the M31 subgroup without And II and EGB0427+63 (large uncertainties).} 
\end {table}
Despite the small number of objects, the probability to obtain this distribution by chance is $(0.07)$.

\paragraph {Remote galaxies subgroup}

The configuration of the remote galaxies can also be compared with the M31 quantization law and with the MW quantization law (see the histograms). 

\begin{table}[!ht]
\begin{center}

\begin{tabular}{cccc}
\includegraphics[width=3cm]{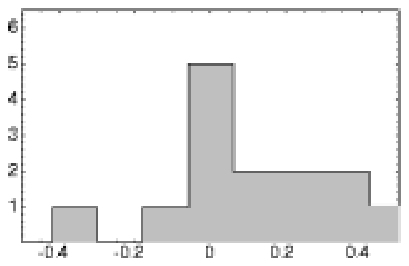} & 
\includegraphics[width=3.1cm]{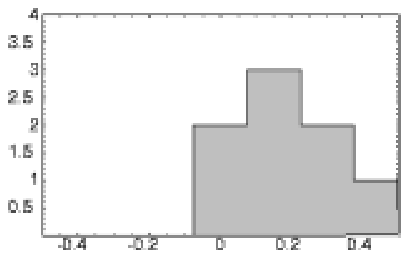} & 
\includegraphics[width=3cm]{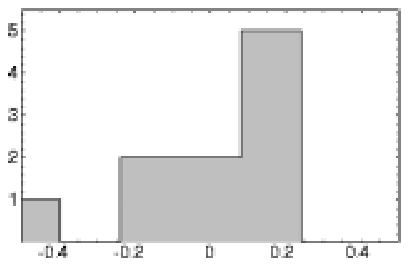} &
\includegraphics[width=3cm]{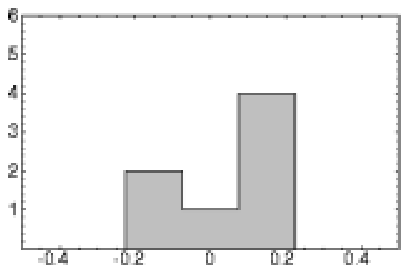} \\
\end{tabular}
\caption {Histograms of the difference between the variable $\widetilde{n}=\sqrt{d / d_{0}}$ and its nearest integer $n$, for the subgroup of remote galaxies.}
\end{center}
\end{table}
The histograms show maxima close to zero as theoretically expected. The galaxy distribution continues to agree with the quantization laws despite the larger distance values and the intervention of both potentials. This opens the possibility that the global solution be a linear 
combination of the solutions found independently for each subgroup (such as in a molecular case). 

\paragraph {Final result for positions}
Combining all the gravitational sub-systems of the Local Group, we 
can now draw a global histogram of the relative differences $\delta n= \widetilde{n}-n$. Since this analysis accumulates all the values tested by the proposed quantization laws, some remote galaxies that agree with both M31 and MW laws are estimated twice, and furthermore, there are some galaxies with large uncertainties. Therefore we also consider a second limited sample: for the whole Local Group, we take 11 galaxies of the Milky Way subgroup and three galaxies of the NGC-3109 subgroup subjected to the quantization law  $d_{n}=5.50 \, n ^ {2}$ kpc.  For M31 ($d_{n}=10.72 \, n ^ {2}$ kpc), we consider only 10 elements of M31 subgroup and 7 remote galaxies (more significant in M31 potential). The histograms obtained for these two samples differ significantly from a uniform distribution and peak at zero as expected. 
\begin{table}[!ht]
\begin {center}
\begin{tabular}{cc}
\includegraphics[width=5.3cm]{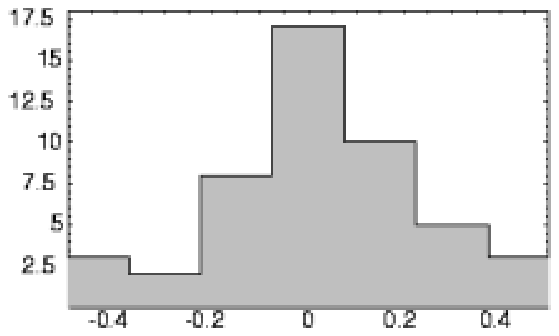} &
\includegraphics[width=5cm]{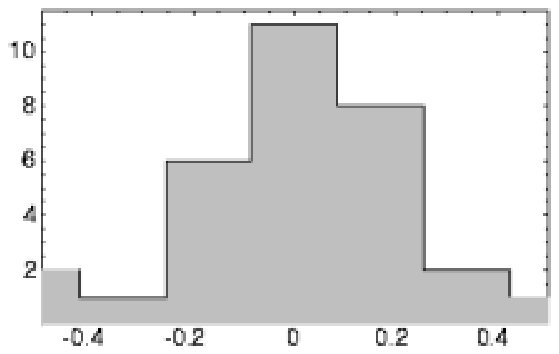} \\
\end{tabular}
\end{center}
\caption{Histograms of the difference between the variable $\widetilde{n}=\sqrt{d / d_{0}}$ and its nearest integer $n$ for all galaxies in the Local Group. The right-hand side histogram is for the limited sample.}
\end {table}

This behavior corresponds to a distribution of $\sqrt{d / d_{0}}$ that shows probability peaks for integer values. The probability to obtain such a distribution by chance is estimated to be $P=2\times 10 ^ {-5} $. This value corresponds to a statistical meaning better than the $4\sigma$ level. This is supported by the limited sample, for which one obtains a similar result, $P=5\times 10^{-5}$.

\subsubsection {Structures in velocity space}

The velocity field of the Local Group provides an interesting test of the theory. We expect the velocity to be quantized according to the formula $< v^ {2} > = (w _ {0} / n) ^ {2}$ (cf. the section about Keplerian dynamical consequences).
The study of the spatial structures has teached us that the constant is $w _ {0}=288$ km/s for the Local Group. A correction by the solar vector is made to obtain galactocentric radial velocities. We use the solar vector defined by Dehen and Binney \cite{binney}. 

A first step of the velocity distribution analysis consists of limiting ourselves to the subset surrounding the Milky Way. We have found that in this case the spatial distribution is given by the law  $r=5.50 \, n ^ {2}$ kpc. The corresponding radial velocity law is given by $ < v ^ {2} >= (w _ {0} / n) ^ {2} $. The solutions in the position space and velocity space are equivalent with regard to the Schr\"odinger equation. Therefore, one associates to a given satellite energy  a position and a velocity state given by the same main quantum number $n$. 

As an example, let us consider the first nearby galaxy (Sagittarius). The distance of this galaxy is characterized by the main number $n=1.7$. From this value we predict a velocity $v=\pm (w _ {0} / 1.7) \approx \pm 168$ km/s, which compares well with the observed galactocentric velocity, $165$ km/s. This example shows the new possibilities offered by our model to 
interpret the peculiar galaxy velocities. 

More generally, one obtains the distance-velocity relation around the MW galaxy in terms of the fundamental constant 288 km/s:
\begin{equation}
\label {vitpos2}
\frac{v}{288 \;{\rm km/s}} =  \pm \sqrt {\frac {5.50 \;{\rm kpc}} {r}} .
\end{equation}

As one can see in Fig.~\ref{fig:phase}, there is a satisfactory agreement of the data with the model based on $w=2 \times 144$ km/s, except for the deviation of some individual galaxies such as Leo I. The right hand side of the figure draws the same distance-velocity diagram completed with the remote galaxies (the Andromeda subgroup is not represented), and matched to the Hubble velocity field at large distances. However, due to the uncertainties on the velocities (which are strongly dependent on the choice of the solar vector) and the small number of objects, we find that the present data in the Local Group does not allow to put to the test a velocity quantization in terms of submultiples of 144 km/s.

\begin{table}
\begin{center}
\begin{tabular}{cc}
\includegraphics[width=7cm]{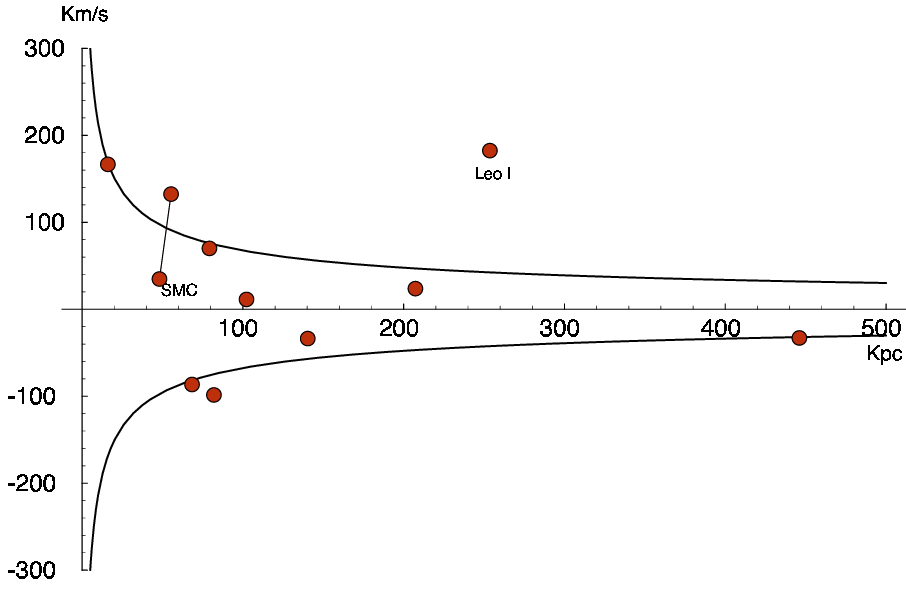} &
\includegraphics[width=7cm]{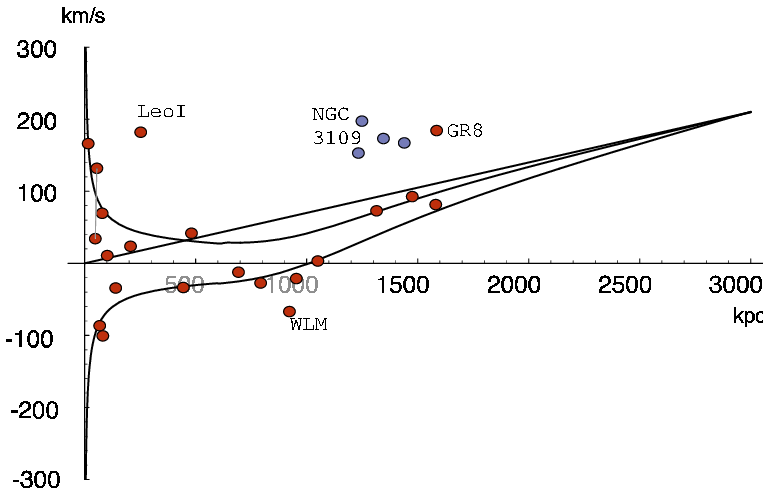} \\
\end{tabular}
\end{center}
\caption{Left hand side: distribution of the galaxy velocities in the MW subgroup in function of their distances. The two curves represent the theoretical prediction from Kepler's third law and a velocity constant of 288 km/s. Right hand side: larger scale representation of the Local Group showing the reconnection of the local Keplerian velocity field to the outer Hubble field.}
\label{fig:phase}
\end{table}

This special study was motivated, despite the small number of objects, by the fact that the Local Group is our own group of galaxies. We intend to extend this work in the future to other similar galaxy groups, which should allow to improve the statistics.

\section{Conclusion and prospect}

We have attempted, in the present contribution, to give an as wide as possible view of the possibilities opened by the scale relativity / gravitational Schr\"odinger approach to the question of the formation and evolution of structures in the Universe.

Despite the fact that observational tests of the theoretical predictions have been suggested from the scale of the Earth to large scale structures of the Universe, we have not presented a fully exhaustive view of the present state of the subject. There are indeed many other astrophysical systems which have not been quoted here, while they come under the same kind of approach and have also began to be analyzed with positive preliminary results: e.g., star radii \cite{lefevre}, the asteroid belt \cite{hermann}, magnetic axes of planets and satellites \cite{galopeau}, the solar wind \cite{galopeau}, etc...

Moreover, we think this contribution should mainly be considered as a working basis, since each of the application domains considered here deserves a special and detailed study. More generally we hope the scale-relativistic approach to be taken as a general tool adapted to finding solutions to various problems of structuration, including in sciences other than physics, particularly in biology \cite{CNG,NCG00,NCG01,synthese2}.

Acknowledgments. We are very grateful to the editors of this special issue for their kind invitation to contribute. We thank Ch. Antoine, P. Galopeau, C. Jacob, Serge Lefranc, Th.Lery, G. Mamon, M. Tricottet, G. Schumacher, N. Tran Minh for helpful discussions, C. Foellmi for suggesting corrections to the manuscript, Sebastien Lefranc and G. Stasinska for communication of useful informations.

\end{document}